\documentclass[times,twocolumn]{aastex631}

\usepackage{savesym}
\savesymbol{tablenum}
\usepackage{siunitx}
\restoresymbol{SIX}{tablenum}

\newcommand{\oiii}{\mbox{[\ion{O}{3}]}}
\newcommand{\hei}{\mbox{\ion{He}{1}}}

\newcommand{\cii}{\mbox{[\ion{C}{2}]}}
\newcommand{\nii}{\mbox{[\ion{N}{2}]}}
\newcommand{\sii}{\mbox{[\ion{S}{2}]}}
\newcommand{\siii}{\mbox{[\ion{S}{3}]}}
\newcommand{\hb}{\mbox{H$\beta$}}
\newcommand{\ha}{\mbox{H$\alpha$}}

\newcommand\lsun{\mbox{\si{L_\odot}}}

\newcommand\mstar{\mbox{$M_\mathrm{star}$}}

\renewcommand{\micron}{\si{\micro\meter}}
\newcommand{\zsp}{\mbox{$z_\mathrm{spec}$}}
\newcommand{\zph}{\mbox{$z_\mathrm{phot}$}}
\newcommand{\lir}{\mbox{$L_\mathrm{IR}$}}

\newcommand{\textred}[1]{\textcolor{red}{#1}}

\renewcommand{\textred}[1]{{#1}}

\received{August 26, 2024}
\revised{November 17, 2024}
\accepted{December 8, 2024}

\submitjournal{ApJ}

\begin{document}

\title{A SPectroscopic survey of biased halos In the Reionization Era (ASPIRE): \\ Spectroscopically Complete Census of Obscured Cosmic Star Formation Rate Density at $z=4-6$}

\correspondingauthor{Fengwu Sun}
\email{fengwu.sun@cfa.harvard.edu}


\suppressAffiliations

\author[0000-0002-4622-6617]{Fengwu Sun}
\affiliation{Center for Astrophysics $|$ Harvard \& Smithsonian, 60 Garden St., Cambridge, MA 02138, USA}

\author[0000-0002-7633-431X]{Feige Wang}
\affiliation{Department of Astronomy, University of Michigan, 1085 S. University Ave., Ann Arbor, MI 48109, USA}
\affiliation{Steward Observatory, University of Arizona, 933 N Cherry Avenue, Tucson, AZ 85721, USA}

\author[0000-0001-5287-4242]{Jinyi Yang}
\affiliation{Department of Astronomy, University of Michigan, 1085 S. University Ave., Ann Arbor, MI 48109, USA}
\affiliation{Steward Observatory, University of Arizona, 933 N Cherry Avenue, Tucson, AZ 85721, USA}

\author[0000-0002-6184-9097]{Jaclyn B. Champagne}
\affiliation{Steward Observatory, University of Arizona, 933 N Cherry Avenue, Tucson, AZ 85721, USA}

\author[0000-0002-2662-8803]{Roberto  Decarli}
\affiliation{INAF–Osservatorio di Astrofisica e Scienza dello Spazio, via Gobetti 93/3, I-40129, Bologna, Italy}

\author[0000-0003-3310-0131]{Xiaohui Fan}
\affiliation{Steward Observatory, University of Arizona, 933 N Cherry Avenue, Tucson, AZ 85721, USA}

\author[0000-0002-2931-7824]{Eduardo Ba{\~n}ados}
\affiliation{Max Planck Institut f\"ur Astronomie, K\"onigstuhl 17, D-69117, Heidelberg, Germany}

\author[0000-0001-8467-6478]{Zheng Cai}
\affiliation{Department of Astronomy, Tsinghua University, Beijing 100084, China}

\author[0000-0002-9090-4227]{Luis Colina}
\affiliation{Centro de Astrobiolog\'{\i}a (CAB), CSIC-INTA, Ctra. de 
Ajalvir km 4, Torrej\'on de Ardoz, E-28850, Madrid, Spain}

\author[0000-0003-1344-9475]{Eiichi Egami}
\affiliation{Steward Observatory, University of Arizona, 933 N Cherry Avenue, Tucson, AZ 85721, USA}

\author[0000-0002-7054-4332]{Joseph F. Hennawi}
\affiliation{Department of Physics, University of California, Santa Barbara, CA 93106-9530, USA}
\affiliation{Leiden Observatory, Leiden University, Niels Bohrweg 2, NL-2333 CA Leiden, Netherlands}

\author[0000-0002-5768-738X]{Xiangyu Jin}
\affiliation{Steward Observatory, University of Arizona, 933 N Cherry Avenue, Tucson, AZ 85721, USA}

\author[0000-0003-1470-5901]{Hyunsung D. Jun}
\affiliation{Department of Physics, Northwestern College, 101 7th St SW, Orange City, IA 51041, USA}
\affiliation{School of Physics, Korea Institute for Advanced Study, 85 Hoegiro, Dongdaemun-gu, Seoul 02455, Republic of Korea}

\author[0000-0002-7220-397X]{Yana Khusanova}
\affiliation{Max-Planck-Institut f\"{u}r Astronomie, K\"{o}nigstuhl 17, D-69117 Heidelberg, Germany}

\author[0000-0001-6251-649X]{Mingyu Li}
\affiliation{Department of Astronomy, Tsinghua University, Beijing 100084, China}

\author[0000-0001-5951-459X]{Zihao Li}
\affiliation{Cosmic Dawn Center (DAWN), Denmark}
\affiliation{Niels Bohr Institute, University of Copenhagen, Jagtvej 128, DK-2200 Copenhagen N, Denmark}

\author[0000-0001-6052-4234]{Xiaojing Lin}
\affiliation{Department of Astronomy, Tsinghua University, Beijing 100084, China}
\affiliation{Steward Observatory, University of Arizona, 933 N Cherry Avenue, Tucson, AZ 85721, USA}

\author[0000-0003-3762-7344]{Weizhe Liu
}
\affiliation{Steward Observatory, University of Arizona, 933 N Cherry Avenue, Tucson, AZ 85721, USA}

\author[0000-0001-5492-4522]{Romain A. Meyer}
\affiliation{Department of Astronomy, University of Geneva, Chemin Pegasi 51, 1290 Versoix, Switzerland}

\author[0000-0003-4924-5941]{Maria A. Pudoka}
\affiliation{Steward Observatory, University of Arizona, 933 N Cherry Avenue, Tucson, AZ 85721, USA}

\author[0000-0003-2303-6519]{George H. Rieke}
\affiliation{Steward Observatory, University of Arizona, 933 N Cherry Avenue, Tucson, AZ 85721, USA}

\author[0000-0003-1659-7035]{Yue Shen}
\affiliation{Department of Astronomy, University of Illinois at Urbana-Champaign, Urbana, IL 61801, USA}
\affiliation{National Center for Supercomputing Applications, University of Illinois at Urbana-Champaign, Urbana, IL 61801, USA}

\author[0000-0003-0747-1780]{Wei Leong Tee}
\affiliation{Steward Observatory, University of Arizona, 933 N Cherry Avenue, Tucson, AZ 85721, USA}

\author[0000-0001-9024-8322]{Bram Venemans}
\affiliation{Leiden Observatory, Leiden University, Niels Bohrweg 2, NL-2333 CA Leiden, Netherlands}

\author[0000-0003-4793-7880]{Fabian Walter}
\affiliation{Max Planck Institute for Astronomy, K\"onigstuhl 17, D-69117 Heidelberg, Germany}

\author[0000-0003-0111-8249]{Yunjing Wu}
\affiliation{Department of Astronomy, Tsinghua University, Beijing 100084, China}

\author[0000-0002-0123-9246]{Huanian Zhang}
\affiliation{Department of Astronomy, Huazhong University of Science and Technology, Wuhan, Hubei 430074, China}

\author[0000-0002-3983-6484]{Siwei Zou}
\affiliation{Chinese Academy of Sciences South America Center for Astronomy, National Astronomical Observatories, CAS, Beijing 100101, China}
\affiliation{Department of Astronomy, Tsinghua University, Beijing 100084, China}

\begin{abstract}

We present a stringent measurement of the dust-obscured star-formation rate density (SFRD) at $z=4-6$ from the ASPIRE JWST Cycle-1 medium and ALMA Cycle-9 large program.
We obtained JWST/NIRCam grism spectroscopy and ALMA 1.2-mm continuum map along 25 independent quasar sightlines, covering a total survey area of $\sim$35\,arcmin$^2$ where we search for dusty star-forming galaxies (DSFGs) at $z = 0 - 7$.
We identify eight DSFGs in seven fields at $z=4-6$ through the detection of \ha\ or \oiii\,$\lambda$5008 lines, including fainter lines such as \hb, \oiii\,$\lambda$4960, \nii\,$\lambda$6585, \sii\,$\lambda\lambda$6718,6733 for six sources.
With this spectroscopically complete DSFG sample at $z=4-6$ and negligible impact from cosmic variance (shot noise), we measure the infrared luminosity function (IRLF) down to $L_\mathrm{IR} \sim 2\times10^{11}$\,\lsun.
We find flattening of IRLF at $z=4-6$ towards the faint end (power-law slope $\alpha = 0.59_{-0.45}^{+0.39}$).
We determine the dust-obscured cosmic SFRD at this epoch as $\log[\rho_\mathrm{SFR,IR} / (\mathrm{M}_\odot\,\mathrm{yr}^{-1}\,\mathrm{Mpc}^{-3})] = -1.52_{-0.13}^{+0.14}$.
This is significantly higher than previous determination using ALMA data in the Hubble Ultra Deep Field, which is void of DSFGs at $z=4-6$ because of strong cosmic variance (shot noise).
We conclude that the majority ($66\pm7$\%) of cosmic star formation at $z \sim 5$ is still obscured by dust.
We also discuss the uncertainty of SFRD propagated from far-IR spectral energy distribution and IRLF at the bright end, which will need to be resolved with future ALMA and JWST observations.

\end{abstract}


\section{Introduction}
\label{sec:01_intro}

It is now clear that the cosmic star formation history peaked at around a redshift of two (see review by \citealt{md14}).
Before this so-called ``cosmic noon'', a decline of unobscured cosmic star formation rate densities (SFRD) toward higher redshifts has been suggested by Hubble Space Telescope (HST) observations at rest-frame ultraviolet (UV) wavelength \citep[e.g.,][]{bouwens15, bouwens20, finkelstein15}. 
However, the measurements of dust-obscured SFRD at infrared (IR) to millimeter wavelengths were much more challenging and uncertain.
Although exciting constraints have been placed with multiple telescopes including Herschel \citep[e.g.,][]{burgarella13, gurppioni13, rr16, liu18, wangl19}, James Clerk Maxwell Telescope (JCMT; e.g., \citealt{koprowski17,lim20}) and Atacama Large Millimeter/submillimeter Array \citep[ALMA; e.g.,][]{dunlop17, bouwens20,gruppioni20,khusanova21,zavala21,fujimoto24,magnelli24,traina24}, most of existing studies rely on a long list of assumptions, including the shape of infrared luminosity function (IRLF), mid-to-far-IR spectral energy distribution (SED), and also photometric redshifts that are strongly degenerate with the stellar age and dust attenuation.

One of the most critical uncertainties is from photometric redshifts (\zph).
The vast majority of dusty star-forming galaxies (DSFGs) are found at $z\simeq 2 - 3$ \citep[e.g., see reviews by][]{casey14,hodge20}. Therefore, it is usually very challenging to differentiate true DSFGs at $z>4$ from lower-redshift ``interlopers'' with large \zph\ error (e.g., see the recent dispute of COSBO-7 at \zph\,$>$\,7 but spectroscopic redshift \zsp\,$=$\,2.625; \citealt{lingc24}, \citealt{jins24}).
Many of these DSFGs at \zph\,$\gtrsim$\,4 are totally dust-obscured at HST wavelengths, and therefore also known as ``HST-dark'', ``NIR-dark'', ``$H$-dropout'' or ``$H$-faint'' galaxies \citep[e.g.,][]{huang11, dacunha15, simpson15, fujimoto16, franco18, alcalde19, wangt19, yamaguchi19, sun21b, talia21, enia22, gomez22, kokorev22, manning22, xiao23, tsujita24}, including certain amount of galaxies that are only detected with ALMA \citep[e.g.,][]{decarli17, mazzucchelli19, venemans19, williams19, fudamoto21, wangf23}.
Therefore, it is very challenging to obtain secure redshifts of these galaxies through ground-based optical/near-IR spectroscopy.

At $z \gtrsim 4$, luminous infrared galaxies (LIRG; IR luminosity $10^{11}\,\lsun \leq L_\mathrm{IR} < 10^{12}$\,\lsun) and the bulk of ultra-luminous infrared galaxies (ULIRG; $10^{12}\,\lsun \leq L_\mathrm{IR} < 10^{13}$\,\lsun) are below the Herschel/SPIRE confusion noise limit at 250--500\,\micron\ \citep[e.g.,][]{nguyen10}, making it difficult to tightly constrain redshifts through far-IR SED except for extraordinarily luminous and/or gravitationally lensed galaxies \citep[e.g.,][]{riechers13, rawle14, sun22a}.
Large millimeter interferometers (e.g., ALMA, Plateau de Bure Interferometer, PdBI and its successor, Northern Extended Millimeter Array, NOEMA) can obtain redshifts of distant DSFGs through CO and \cii\,$\lambda$158\,\micron\ line scans.
However, such type of observations are time-consuming for high-redshift LIRGs and ULIRGs (e.g., 8.3-hour on-source for the ALMA CO scan of MAMBO-9 at \zsp\,$=$\,5.85 with \lir\,$\sim$\,$4\times10^{12}$\,\lsun; \citealt{casey19,jin19}), with again exceptions for extraordinarily luminous sources at observed \lir\,$\gtrsim$\,$10^{13}$\,\lsun\ \citep[e.g.,][]{weiss13, strandet16,reuter20,chenc22}.

Without spectroscopic redshifts, it is very challenging to securely select a sample of DSFGs at $z>4$, directly measure their IRLF and accurately determine the obscured SFRD.
For example, based on serendipitous (sub-)millimeter continuum sources discovered by Cycle-5 \textsl{``ALMA Large Program to INvestigate CII at Early Times''} (ALPINE, \citealt{lefevre20}), \citet{gruppioni20} found that the obscured SFRD remains almost constant across $z\simeq 2 - 6$.
This is in great contrast to the results based on Cycle-4 large program \textsl{``ALMA Spectroscopic Survey in the Hubble Ultra Deep Field (HUDF)''} (ASPECS; \citealt{walter16}), where no DSFGs at $z>4$ was blindly discovered with deep 1-mm and 3-mm survey of the HUDF (\citealt{aravena20}; also \citealt{dunlop17}; \citealt{hatsukade18}).
Alternative methods have been developed to infer obscured SFRD through the relation between infrared excess (IRX\,=\,$L_\mathrm{IR} / L_\mathrm{UV}$, i.e., IR-to-UV luminosity ratio) and UV continuum slope $\beta_\mathrm{UV}$ \citep[e.g.,][]{meurer99} or stellar mass \mstar\ \citep[e.g.,][]{bouwens20, khusanova21, algera23, magnelli24}, or a backward modeling approach that could infer the redshift evolution of IRLF (and thus obscured SFRD) through (sub)-millimeter number counts \citep{casey18,casey18b, zavala21}.
Nevertheless, the obscured SFRD at $z\sim 5$ measured based on ASPECS \citep[e.g.,][]{bouwens20,zavala21} data is still lower than that measured with ALPINE by $\sim$\,1\,dex \citep[][]{gruppioni20}.

Most recently, direct measurements through ALMA Cycle-6 large program \textsl{``ALMA Lensing Cluster Survey''} \citep[ALCS;][]{kohno23} suggest that the obscured SFRD at $z\sim5$ is between those inferred through ALPINE and ASPECS programs \citep{fujimoto24}.
Despite significant advance made with ALCS in probing DSFGs towards sub-LIRG-luminosity (\lir\,$<$\,$10^{11}$\,\lsun) and reducing cosmic variance (shot noise) through 33 lensing cluster fields, caution is still necessary as the ALCS sample of DSFGs at $z>4$ is not yet spectroscopically complete.

JWST \citep{gardner23} provides unprecedented opportunities to obtain near-IR spectroscopy of DSFGs at high redshifts.
Specifically, with NIRCam wide-field slitless spectroscopy (WFSS) at 2.4--5.0\,\micron\ \citep{rieke23}, it is now possible to obtain near-IR spectra of all galaxies that enter the field of view (FoV; up to $\sim$\,9\,arcmin$^2$).
It has been clear from JWST Cycle-1 programs that \ha\ lines from HST-dark DSFGs at $z > 5$ can be easily detected with NIRCam WFSS with an on-source integration time of 1--2\,hours \citep[e.g.,][]{hd24, sun24, williams24, xiao24}.
This promises a highly complete spectroscopic survey of dust-obscured star formation history towards the Epoch of Reionization.

In this work, we present a stringent measurement of IRLF and obscured SFRD at $z=4-6$ through the ASPIRE JWST Cycle-1 medium and ALMA Cycle-9 large program (Wang, F.\ et al.\ in prep.).
Through ALMA and JWST spectroscopic survey of DSFGs over 25 independent quasar sightlines, we discovered eight DSFGs at $z_\mathrm{spec} = 4-6$ down to 1.2-mm flux density of $S_\mathrm{1.2mm} \sim 0.15$\,mJy.
We describe the observations and data processing techniques in Section~\ref{sec:02_obs}, present the redshift confirmation and analyses of their physical properties in Section~\ref{sec:03_ana}.
We then study their dust obscuration,  contribution to 1.2--mm number count, IRLF and obscured SFRD in Section~\ref{sec:04_res}.
Further implications and cautions are discussed in Section~\ref{sec:05_dis}.
Our conclusions are summarized in Section~\ref{sec:06_sum}.

Throughout this work, we assume a flat $\Lambda$CDM cosmology with $H_0= 70$\,\si{km.s^{-1}.Mpc^{-1}}, $\Omega_\mathrm{M} = 0.3$ and $\sigma_8 = 0.8$. 
AB magnitude system \citep{oke83} is adopted to describe source brightness in the optical and near-IR.
We also assume a \citet{chabrier03} initial mass function.
We define the IR luminosity ($L_\mathrm{IR}$) as the integrated luminosity over a rest-frame wavelength range from 8 to 1000\,\micron.

\section{Observation and Data Reduction}
\label{sec:02_obs}

\subsection{JWST/NIRCam}
\label{ss:02a_jwst}

A SPectroscopic survey of biased halos In the Reionization Era (ASPIRE) started as a 62-hour JWST Cycle-1 medium program (PI: Wang, F.; Program ID: 2078).
With the powerful NIRCam WFSS observing mode, ASPIRE observed 25 luminous quasars at $z=6.5 - 6.8$ through NIRCam grism WFSS in the F356W filter (3.1--4.0\,\micron) and imaging in the F115W, F200W and F356W filter (1--4\,\micron).
The observational design of ASPIRE JWST program will be presented by Wang, F.\ et al. (in prep.; see also \citealt{wangf23}; \citealt{yangj23}).
The primary science goal of ASPIRE is to detect companion galaxies at the quasar redshifts through \oiii\,$\lambda\lambda$4960,5008 and \hb\ lines, and therefore answer whether the most massive black holes in the early Universe reside in the most massive dark matter halos \citep{wangf23}.
ASPIRE has also enabled a large number of other scientific investigations including supermassive black hole mass measurements \citep{yangj23}, morphology of quasar host galaxies and outflows (Yang, J.\ et al.\ in prep.), the environmental impact on galaxy properties and evolution history at $z\sim 6.5$ \textred{\citep{champagne25a,champagne25b}}, the connection between intergalactic medium (IGM) opacity and galaxies at $z\sim6$ \textred{\citep{jinx24}}, and the metal enrichment of circumgalactic medium (CGM) at $z\simeq 4 - 6$ \citep[][]{wu23b,zou24}.

\begin{figure*}[!t]
\centering
\includegraphics[width=\linewidth]{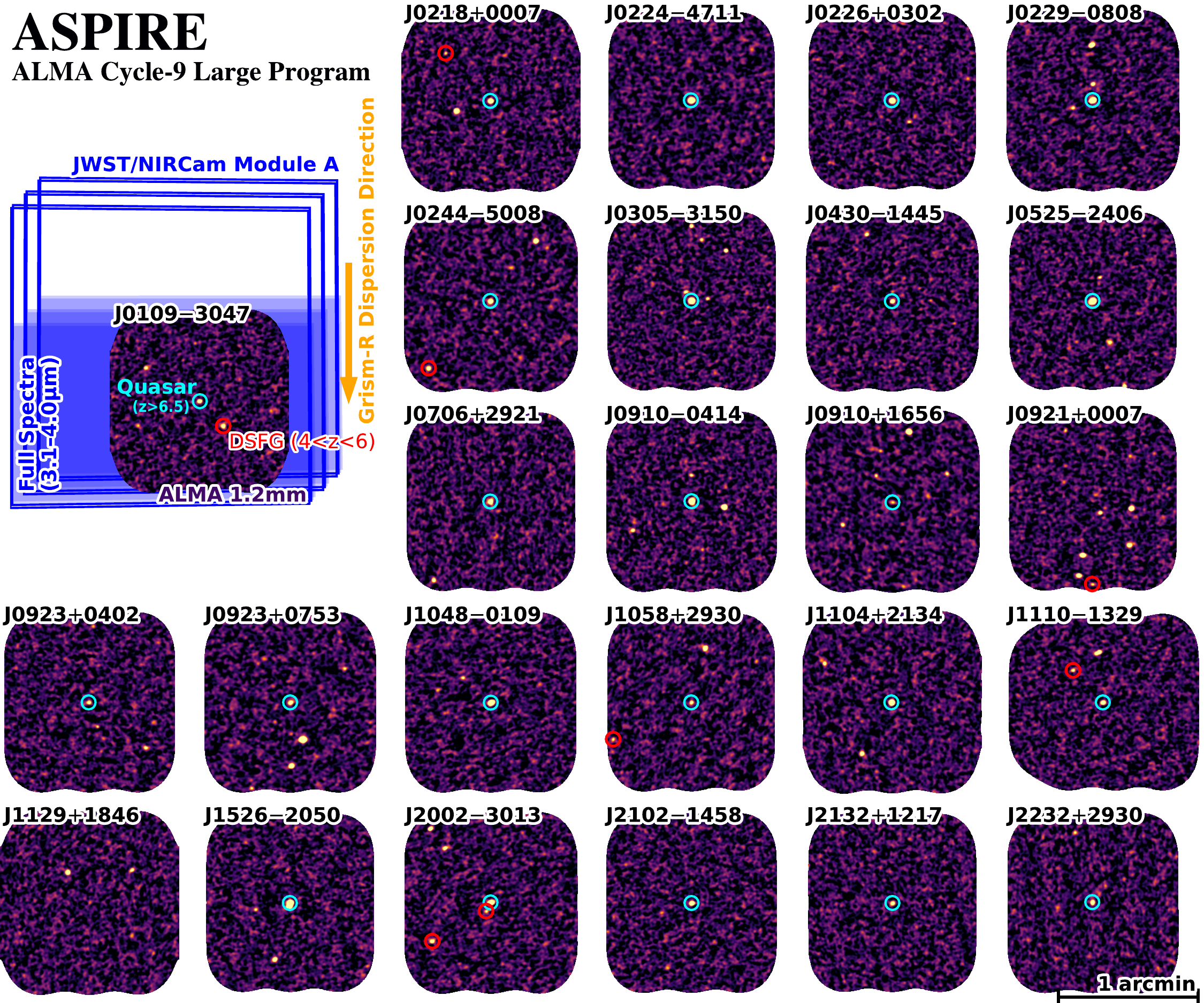}
\caption{1.2-mm continuum images of 25 quasar fields obtained by ASPIRE ALMA Cycle-9 large program. In the top-left panel, we highlight the design of JWST/NIRCam and ALMA observations. 
The whole ALMA 1.2-mm continuum imaging mosaics (\textit{uv}-tapered with FWHM\,=\,1\arcsec) are within the full spectral ($\lambda$\,=\,3.1--4.0\,\micron\ with F356W filter) coverage region of NIRCam module A as indicated by the blue shaded region. 
The quasar J0109--3047 ($z=6.791$; cyan circle) is located in the center of ALMA footprint, and one DSFG (J0109m3047.C02 at $z=5.549$) is also highlighted with red circle. 
The position angle of NIRCam WFSS observation is 270\arcdeg\ for J0109 field, and therefore the grism-R dispersion direction is almost from north to south as indicated by the orange arrow. 
Note that the dispersion direction depends on the JWST/NIRCam PA and varies from field to field.
ALMA 1.2-mm continuum images of all the other 24 fields are also displayed. Quasars with continuum detection ($z=6.5-6.8$) and spectroscopically confirmed DSFGs at $z=4-6$ are highlighted in cyan and red circles, respectively.
Many DSFGs at other redshifts are also detected with ALMA but not highlighted in this figure.
}
\label{fig:all_alma}
\end{figure*}

In all of the 25 quasar fields observed with ASPIRE, the NIRCam WFSS on-source integration time with the F356W filter is 2834\,s. 
Grism R was used, which disperses light along the row (horizontal) direction of the detector at a spectral resolution of $R \simeq 1300 - 1600$. 
The quasar was placed in the full spectral coverage region in module A (see \citealt{wangf23} and Figure~\ref{fig:all_alma}) because of higher throughput \citep{rieke23}.
During the NIRCam WFSS exposure through the long-wavelength (LW) channel, the short-wavelength (SW) channel of NIRCam obtained direct imaging observation with the F200W filter. 
After the WFSS exposure, direct imaging observations were obtained in F115W and F356W filter through SW and LW channel, respectively, and the exposure time is 1417\,s for both filters.

ASPIRE NIRCam data reduction has been described by \citet{wangf23} and \citet{yangj23} and will be described in full details by Wang, F.\ et al.\ (in prep.). Here we only briefly described the key steps adopted in the data processing routine.

The NIRCam imaging data were processed based on \textsc{jwst} calibration pipeline \citep{bushouse_jwst} \verb|v1.10.2| with the reference files \verb|jwst_1080.pmap| (including JWST Cycle-1 NIRCam flux calibrations).
Customized steps including the {\it 1/f} noise subtraction along both row and column directions  on the stage-1 data products (\texttt{\_rate} files).
The sky background was determined iteratively and subtracted after masking detected sources in the images.
The world coordinate systems (WCS) of all images are aligned to Gaia DR3 \citep{gaiadr3} or DESI Legacy Imaging Survey \citep{dey19} if not enough Gaia stars are found within the FoV.
The flux-calibrated, WCS-registered and background-subtracted stage-2 data products  (\texttt{\_cal} files) were then passed through the stage-3 pipeline to create mosaicked images in each band.
The final mosaicked images in each band were produced with fixed pixel size 0\farcs031 (for F115W and F200W; native SW pixel size) and 0\farcs0315 (for F356W; half of native LW pixel size) with \texttt{pixfrac = 0.8}.

The NIRCam WFSS data were processed using the combination of \textsc{jwst} calibration pipeline \verb|v1.10.2| \citep{bushouse_jwst} and customized scripts.
Starting from stage-1 data products, we subtract \textit{1/f} noise pattern along the column direction (i.e., orthogonal to the grism dispersion direction).
The grism images were then flat-field and assigned with WCS information.
We also subtracted master median background models from the processed images.
After that, the astrometric offsets were measured between the simultaneous SW exposure and the fully calibrated F356W image mosaics, and the offsets were applied to the tracing and dispersion models \citep[based on commissioning data;][]{sun23}.
We then extracted 2D spectra of sources from each individual grism exposure, and the data were coadded into 2D spectra resampled into a common wavelength (9.8 $\rm \AA~pixel^{-1}$; i.e., native dispersion) and spatial grids following the histogram2D technique in the \textsc{PypeIt} software \citep{prochaska20}.
We then extracted 1D spectra from the coadded 2D spectra using both optimal \citep{Horne86} and boxcar extraction algorithms.
Because NIRCam WFSS mode effectively obtains spectra of all sources entering its FoV, contaminants are often seen in 2D and 1D spectra.
In this work, we modeled and subtracted the continuum emission (mostly from contaminants) in 1D spectra through spline interpolation after masking the emission line wavelength range ($|\Delta \lambda_\mathrm{obs}| \leq 0.05$\,\micron).
We also visually inspected 2D, 1D spectra and JWST images to identify and reject emission-line contaminants.

\subsection{ALMA}
\label{ss:02b_alma}

ALMA 1.2-mm mosaics of all 25 ASPIRE quasar fields were obtained through a 100-hour ALMA Cycle-9 large program (PI: Wang, F.; Program ID: 2022.1.01077.L).
In each of the ALMA band-6 mosaics, we tuned the central frequency of the lower sideband as the \cii\ line frequency at the quasar redshift (min--median--max: 243.0--247.9--252.9\,GHz).
Two 1.875--GHz spectral windows were placed to cover the --2200\,$\sim$\,+2200\,\si{km.s^{-1}} velocity range around the quasar redshift, with the primary goal to blindly detect \cii\ emitters at quasar vicinity and measure the Mpc-scale clustering of \cii\ emitters around these quasars.
Two more spectral windows were placed in the upper sideband to increase the continuum sensitivity, and the final effective frequencies of continuum images are 250.6--255.6--260.6\,GHz (min--median--max).
As illustrated in Figure~\ref{fig:all_alma}, in each quasar field, we map the $\sim$\,1\farcm2$\times$1\farcm1 region centered on the quasar with $\sim$\,23 Nyquist-sampled ALMA 12-m pointings.
The ALMA footprints are covered by the full spectral coverage region of NIRCam WFSS at 3.1--4.0\,\micron, allowing us to search for emission lines from these ALMA sources.

ASPIRE-ALMA observations were conducted from October 14, 2022 to February 14, 2023 through C-2/3 configuration. 
The typical angular resolution of ALMA observation is $\sim$\,0\farcs65.
The total survey area of the ASPIRE-ALMA program is 34.9\,arcmin$^2$ at above a primary beam response limit of 0.25 as adopted in this paper.

We reduced all ALMA data with the standard \textsc{CASA} \verb|v6.4.1.12| pipeline \citep{casa07, casa22} for Cycle-9 data.
Our data reduction started from raw ALMA science data models with the restoration of pipeline-calibrated, observatory-flagged data through \texttt{scriptForPI.py}.
These calibrated measurement sets were then concatenated and imaged based on the North American ALMA imaging script template\footnote{\href{https://github.com/aakepley/ALMAImagingScript}{https://github.com/aakepley/ALMAImagingScript}}.
To avoid the artificial boost of continuum flux densities from \cii\ emitters at quasar redshifts, we first flagged the spectral channels that have a velocity offset smaller than 500\,\si{km.s^{-1}} from the quasar \cii\ line center.
Noisy spectral channels caused by telluric absorption were also visually identified through command \texttt{plotms} and flagged from continuum imaging.
We then split the measurement sets to a channel width of 125\,MHz and obtained continuum imaging at both native ALMA resolution (Briggs weighting \texttt{robust=0.5}, no \textit{uv}--tapering) and tapered resolution (\texttt{robust=2.0}, \textit{uv}--tapered with Gaussian kernel of FWHM\,=\,1\farcs0).
\textred{The reduction with \texttt{robust=0.5} weighting is to optimize the sensitivity for compact emission, and our trail suggests that the sensitivity is similar to that of naturally weighted (\texttt{robust=2.0}, no \textit{uv}--tapering) imaging data products.}
At tapered resolution, sources with extended structural profile (FWHM\,$\gtrsim$\,1\arcsec) can be detected at higher signal-to-noise ratio (S/N) than that on native-resolution image.
The synthesized beam FWHM is $0\farcs65\pm0\farcs05$ and $1\farcs34\pm0\farcs05$ for native and tapered continuum image mosaics, respectively.
The image mosaics were produced with \texttt{tclean} command with a pixel size of 0\farcs12.
We adopted the automatic multi-threshold masking algorithm \citep{kepley20} to save human labor for interactive masking, which is found to be comparably accurate as manual-masking results in our trials.
The continuum root-mean-square (RMS) noises (before primary beam response correction) of our ALMA mosaics are 0.031$\pm$0.004\,mJy\,beam$^{-1}$ and 0.034$\pm$0.004\,mJy\,beam$^{-1}$ at native and tapered resolution, respectively.

ALMA continuum sources were identified on both native and tapered images using the simple \texttt{find\_peaks} algorithm in \textsc{photutils} \citep{photutils}.
The detailed statistics of ALMA continuum sources at all redshifts will be presented by a forthcoming paper from the collaboration.
We select ALMA continuum sources at signal-to-noise ratio $\mathrm{S/N} \geq 5$ in native images or $\mathrm{S/N} \geq 4$ in tapered images (i.e., similar to other ALMA Band-6 continuum surveys, e.g., \citealt{fujimoto24}), resulting in 138 non-repeated peaks.
From all ALMA images, we find that the most significant negative peak is at $5.01\sigma$ in native images and $5.04\sigma$ in tapered images, and above our detection threshold there are 1 and 20 negative peaks in the native \textred{($\mathrm{S/N}\leq-5$)} and tapered \textred{($\mathrm{S/N}\leq-4$)} images, respectively.
Therefore, we visually inspected the JWST images of positive peaks that are less significant than the strongest negative peaks, and rejected 20 positive peaks at $\mathrm{S/N}=4 - 5$ that do not show clear JWST counterpart within 0\farcs7.
The number of rejected positive peaks is the same as the number of negative peaks above our detection threshold, indicating that spurious continuum sources have been properly removed from our sample.

The final ASPIRE ALMA continuum source sample includes 117 sources (primary beam response $\geq$\,0.25) down to a flux density of $S_\mathrm{1.2mm} = 0.11$\,mJy (median: 0.52\,mJy).
Among them, 23 sources are targeted quasars at $z=6.5 - 6.8$, while quasar J0921+0007 ($z=6.5646$; \citealt{yangj21}) and the radio-loud quasar J1129+1846 ($z=6.824$; \citealt{banados21}) are below the ALMA continuum detection limit as shown in Figure~\ref{fig:all_alma}.
After inspecting the JWST images of the remaining 94 sources, we classify all of them as DSFGs at cosmological distances.

\section{Analyses}
\label{sec:03_ana}

\begin{figure*}
\centering
\includegraphics[width=\linewidth]{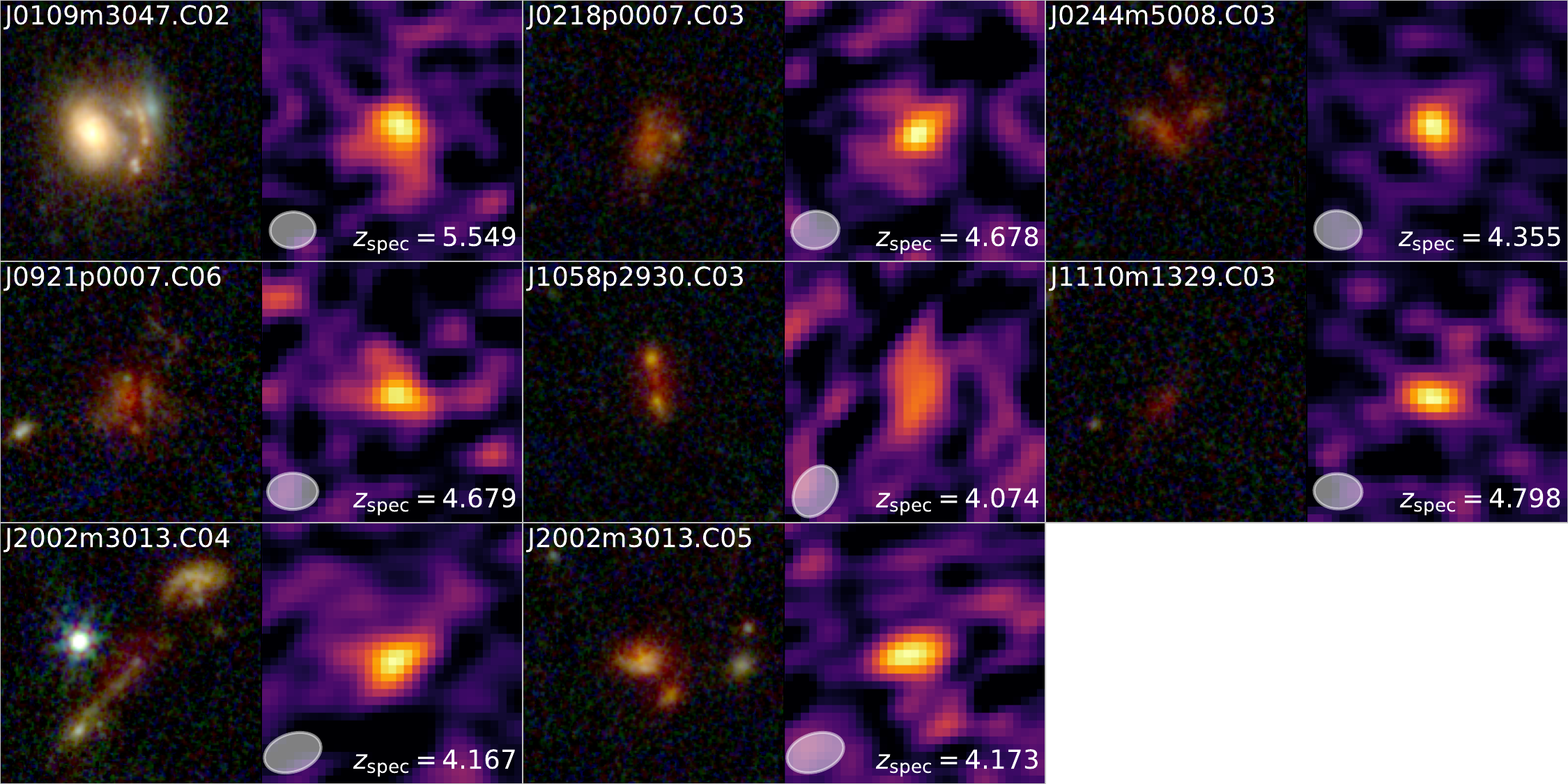}
\caption{JWST NIRCam (red: F356W; green: F200W, blue: F115W) and ALMA 1.2-mm continuum images of DSFGs at $z=4-6$ discovered with the ASPIRE survey. Image sizes are 4\arcsec$\times$4\arcsec\ (north up, east left).
Source ID, spectroscopic redshifts and ALMA beam sizes are indicated in the plots. 
Most sources appear red in JWST RGB images, indicating that they are highly dust-obscured galaxies at high redshifts.
Note that J0109m3047.C02 is gravitationally lensed by the bright galaxy on the left (see Appendix~\ref{apd:01_lens}).
}
\label{fig:cutout}
\end{figure*}

\subsection{Spectroscopic Identification of DSFGs at $z=4-6$}
\label{ss:03a_spec}

To obtain spectroscopic redshifts of the aforementioned 94 ALMA continuum sources (excluding targeted quasars), we extract the NIRCam grism spectra of their JWST counterparts.
At 3.1--4.0\,\micron, we expect to detect emission lines including Paschen\,$\alpha$ (Pa$\alpha$) at $z = 0.7 - 1.1$, Pa$\beta$ at $z = 1.5 - 2.1$, Pa$\gamma$ and \hei\,$\lambda$10833 at $z = 1.9 - 2.7$, \siii\,$\lambda$9533 at $z = 2.3 - 3.1$, \siii\,$\lambda$9071 at $z = 2.5 - 3.5$, and most relevantly, \ha\ at $z = 3.8 - 5.0$ and \oiii\,$\lambda$5008 at $z = 5.3 - 6.9$.

We visually inspect the 2D and 1D grism spectra of all ALMA continuum sources, both before and after continuum subtraction, to identify potential emission lines.
Our visual inspection, together with the ALMA \cii\ line search (will be presented in a forthcoming paper from the collaboration), yields 16 redshifts at $z = 3.8 - 6.8$ through \ha, \oiii\ or \cii\ line detections.
Six sources are at $z>6$, and all of them are quasar companions ($z=6.5-6.8$).
At our interested redshift range $z = 4 - 6$ that is not affected by potential galaxy overdensities associated with quasar, we obtain redshifts for eight sources.

The JWST and ALMA images of these eight sources are displayed in Figure~\ref{fig:cutout}.
All of these sources are highly secure detections in ALMA (S/N\,$\geq$\,6.5), and all of them have JWST counterparts detected in the F200W and F356W band.
The relative astrometric error between JWST and ALMA images is $\lesssim 0\farcs1$.
Most of these sources appear red (i.e., large F200W--F356W and F115W--F356W color) in the JWST RGB image, indicating that they are highly dust-obscured galaxies at high redshifts.
These sources are also clearly resolved at the JWST wavelengths without showing bright point-like structure, indicating that they unlikely host unobscured active galactic nuclei (AGN).
Detailed morphological study of ASPIRE DSFGs will be a focus of future work from the collaboration.

\begin{figure*}[!th]
\centering
\includegraphics[width=0.49\linewidth]{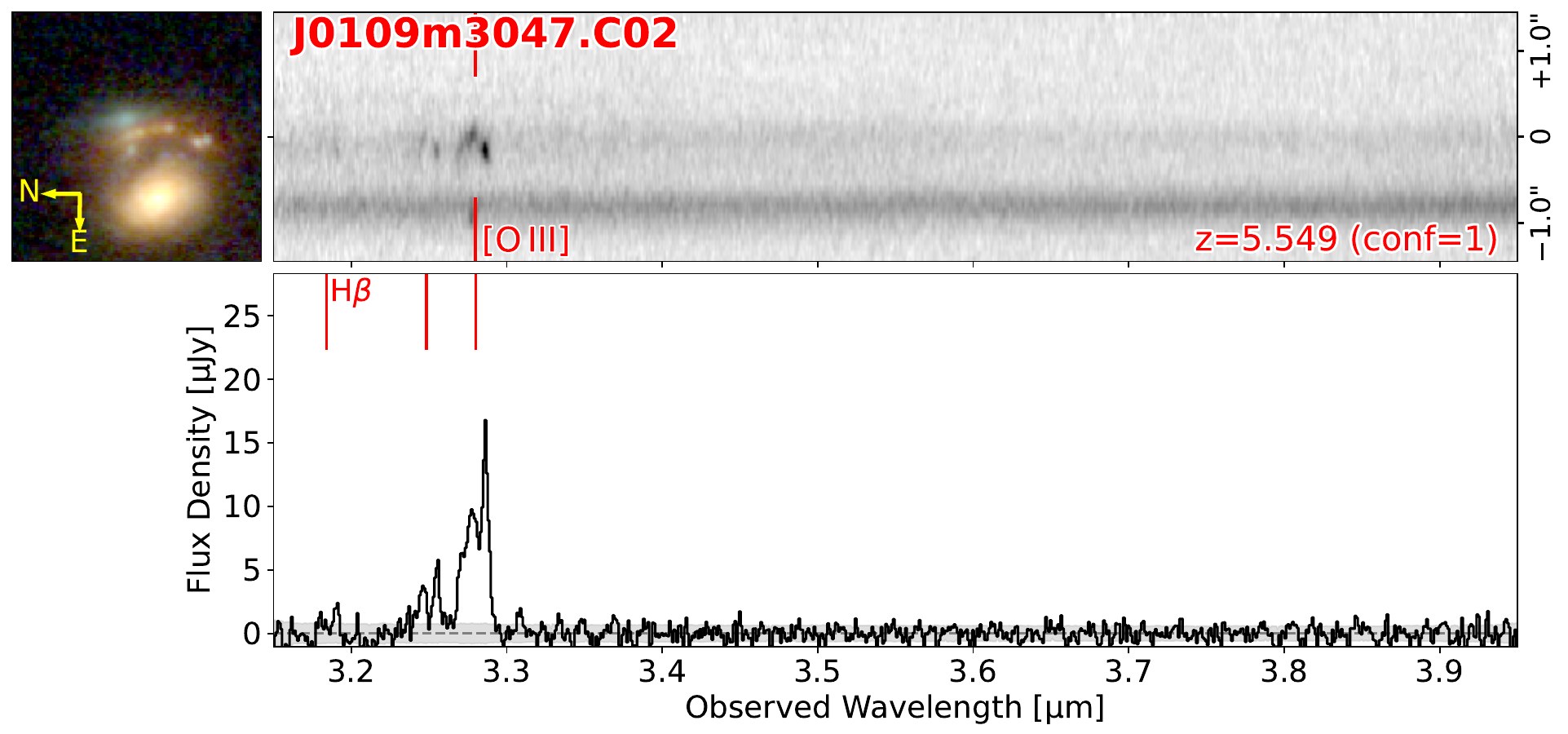}
\includegraphics[width=0.49\linewidth]{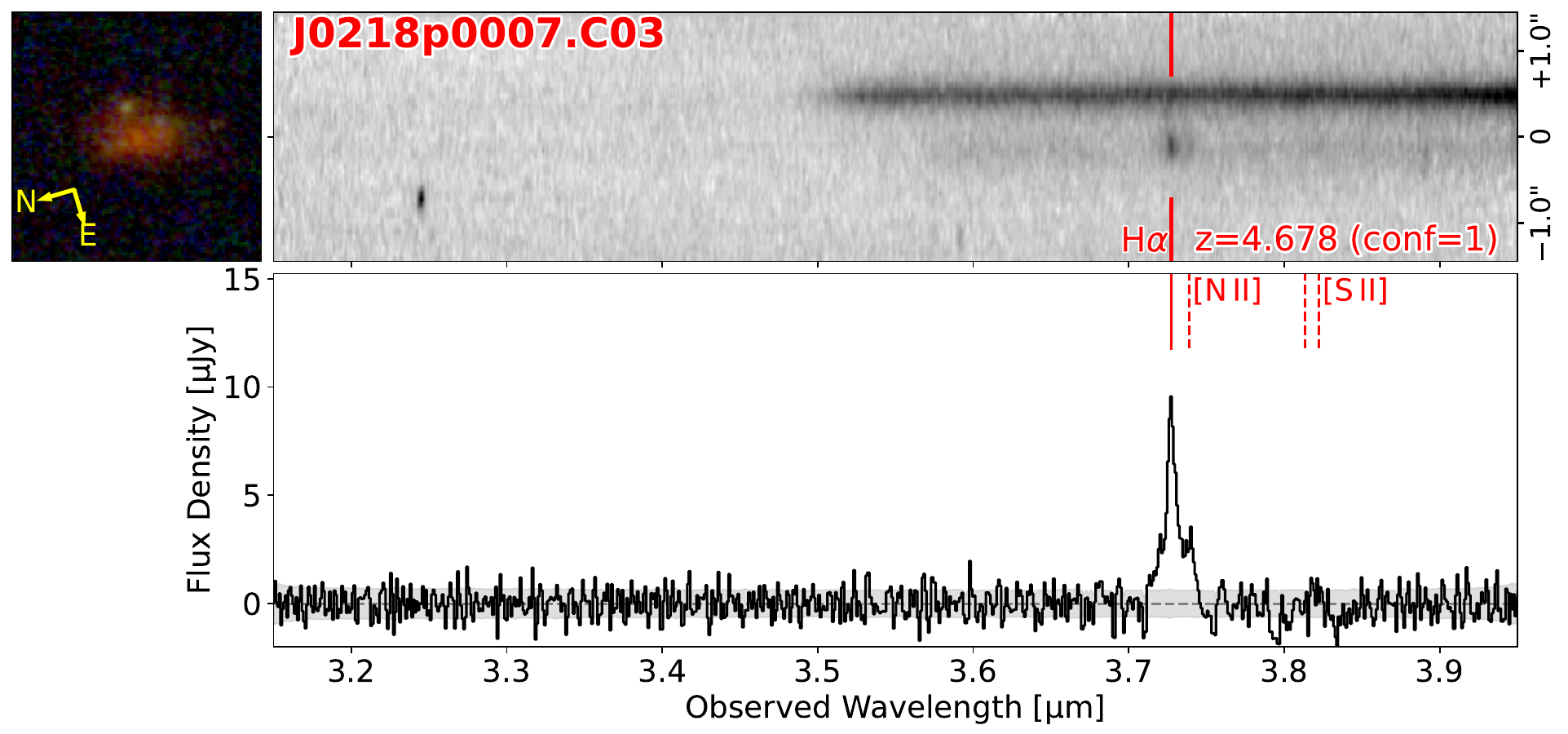}
\includegraphics[width=0.49\linewidth]{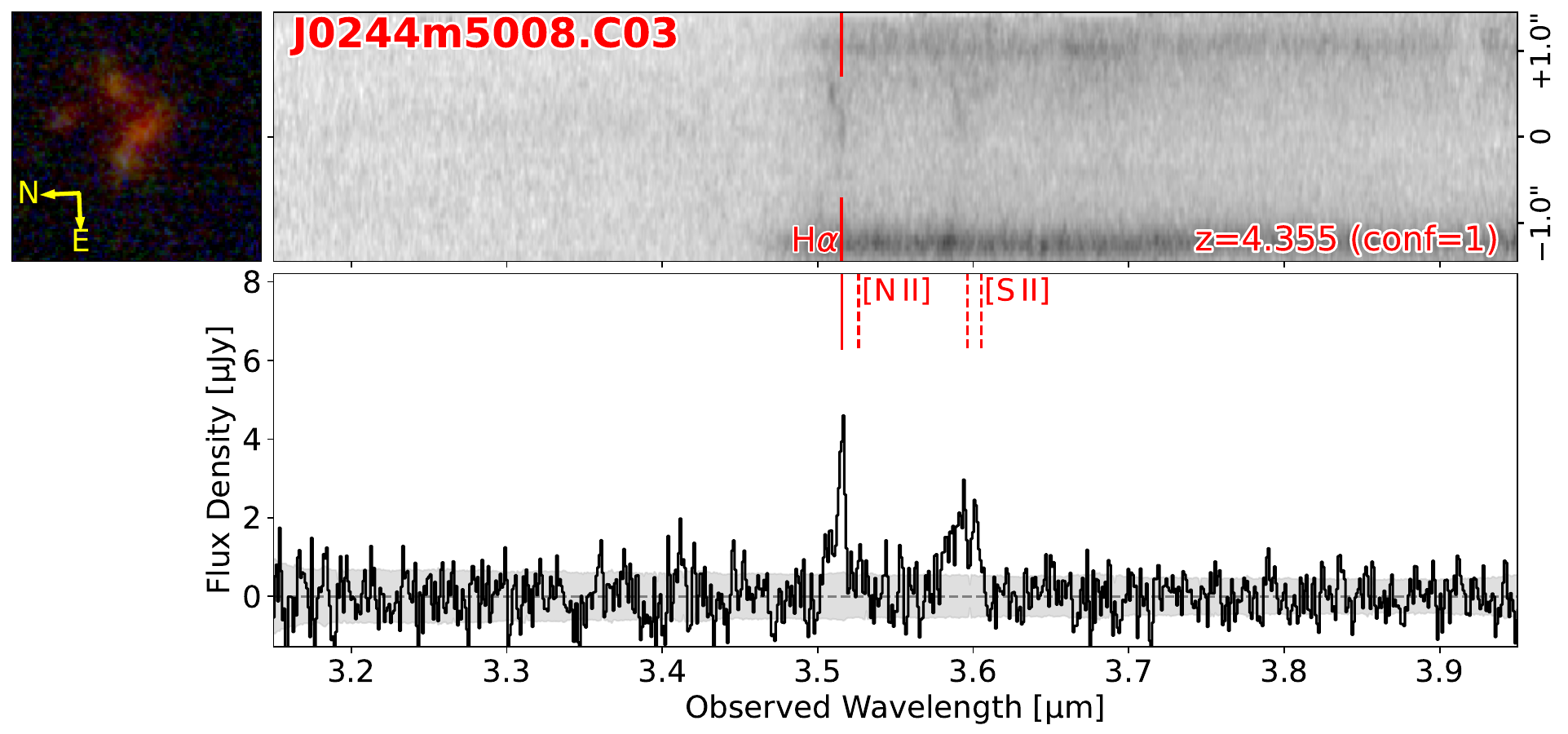}
\includegraphics[width=0.49\linewidth]{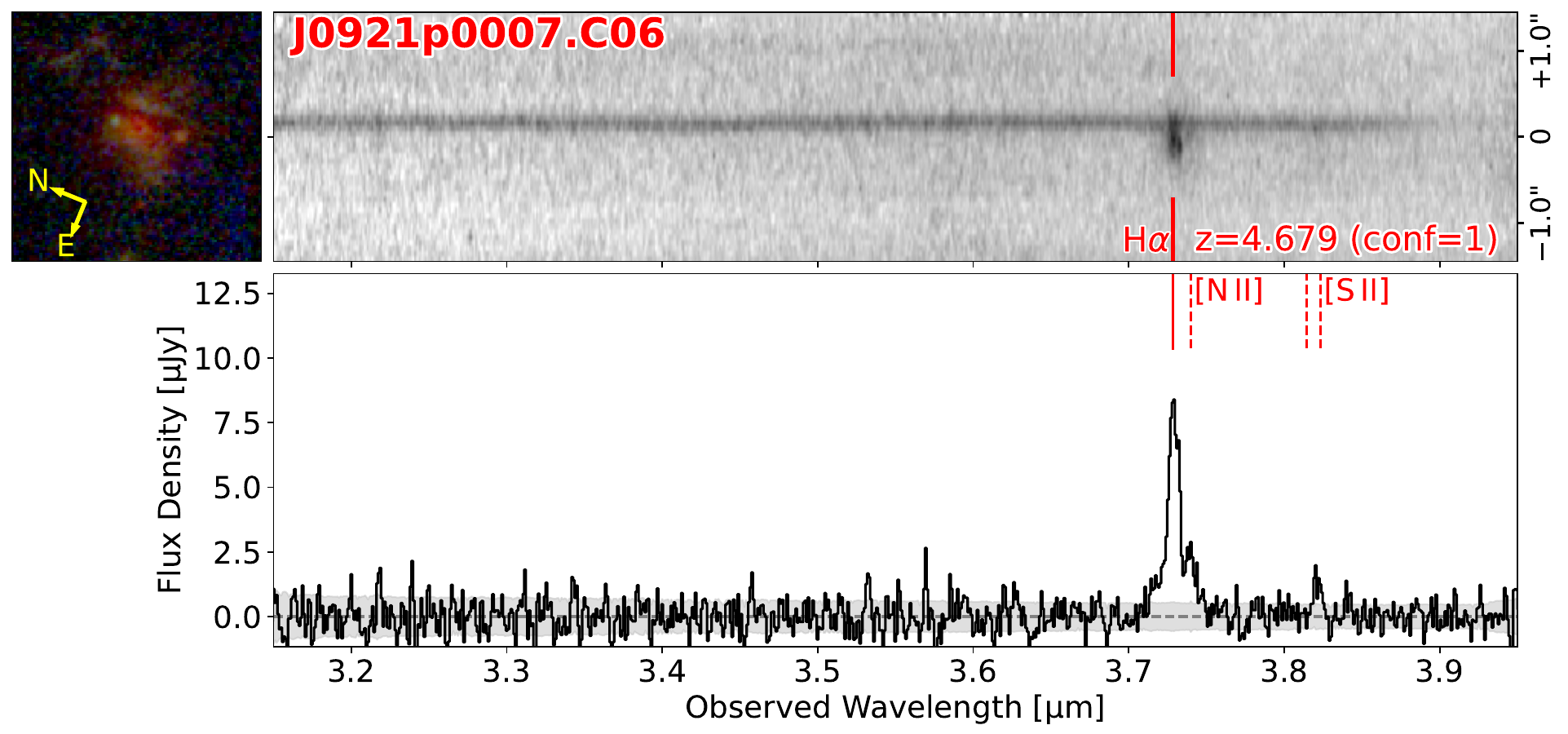}
\includegraphics[width=0.49\linewidth]{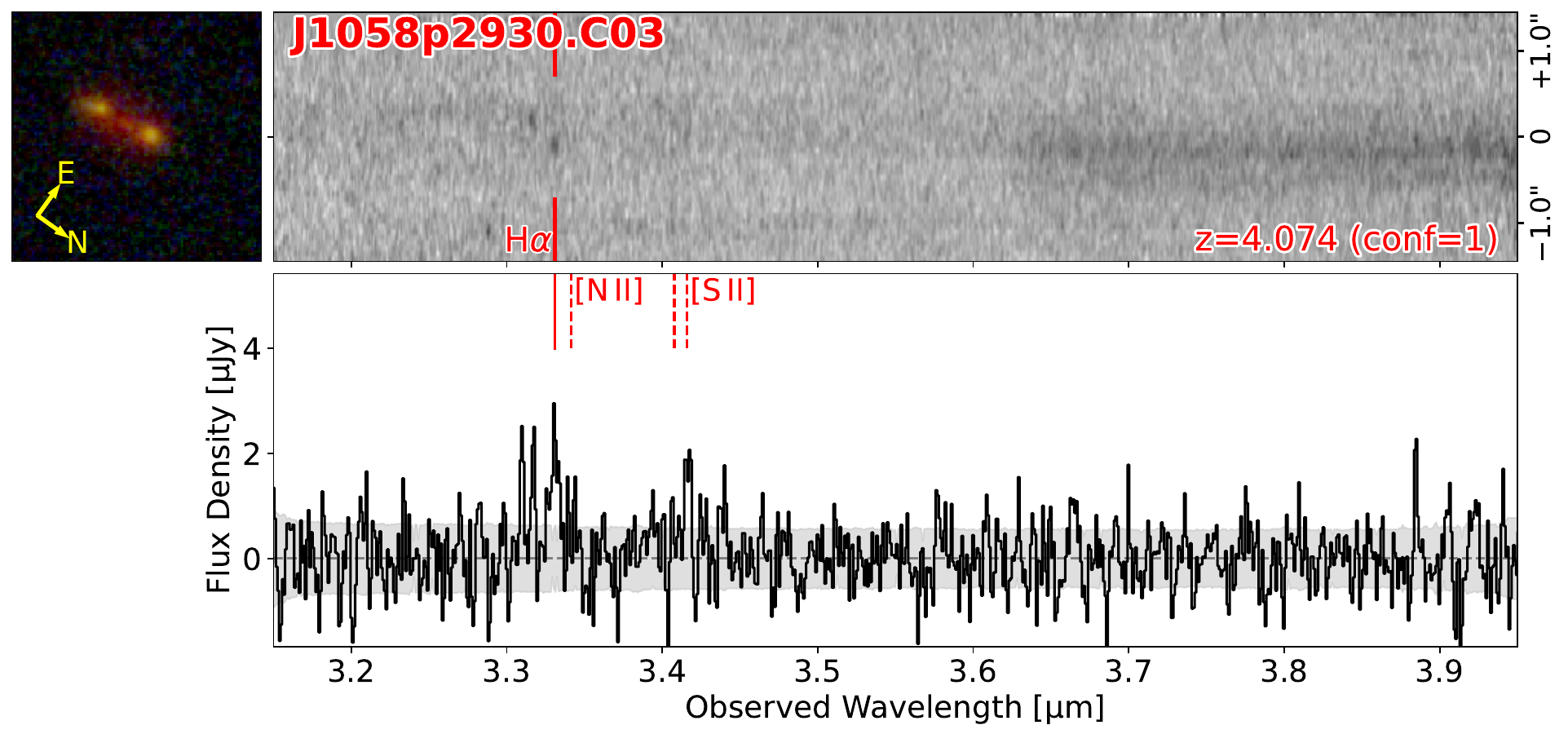}
\includegraphics[width=0.49\linewidth]{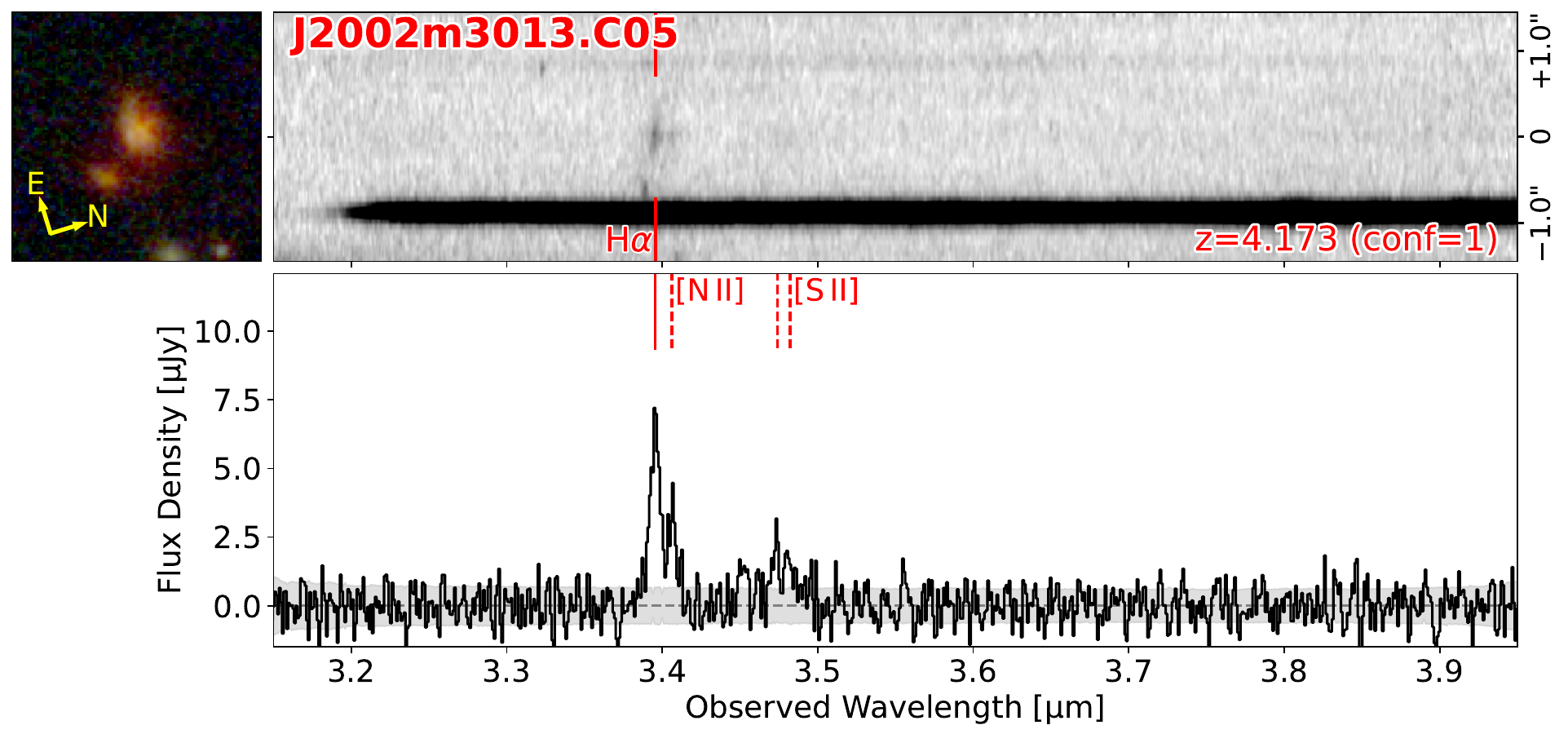}
\caption{JWST NIRCam image, 2D and 1D spectra of six DSFGs with secure spectroscopic redshifts at $4-6$ (\texttt{conf=1}).
Note that the continuum emission (primarily from bright contaminating galaxies) is subtracted in 1D spectra, but not in 2D spectra.
NIRCam images are aligned along the dispersion direction (from left to right).
Also note that the 2D spectra are compressed in the dispersion direction for display purpose.
Primary emission lines (\hb, \oiii\,$\lambda\lambda$4960,5008, \ha) are indicated with solid red lines, and other fainter lines (\nii\,$\lambda$6585, \sii\,$\lambda\lambda$6718,6733) are indicated with dashed red lines. 
}
\label{fig:spec_conf1}
\end{figure*}

\begin{figure*}[!th]
\centering
\includegraphics[width=0.49\linewidth]{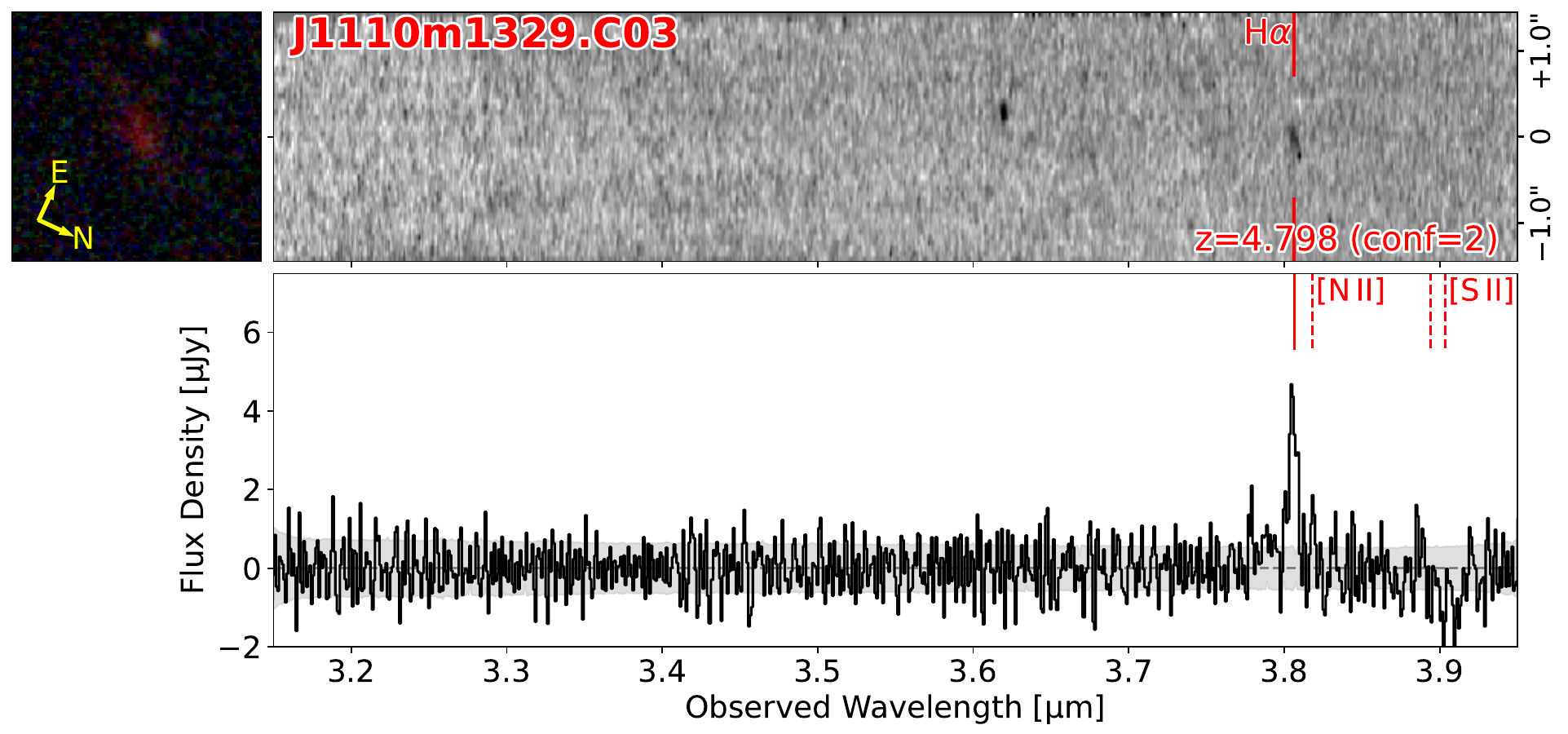}
\includegraphics[width=0.49\linewidth]{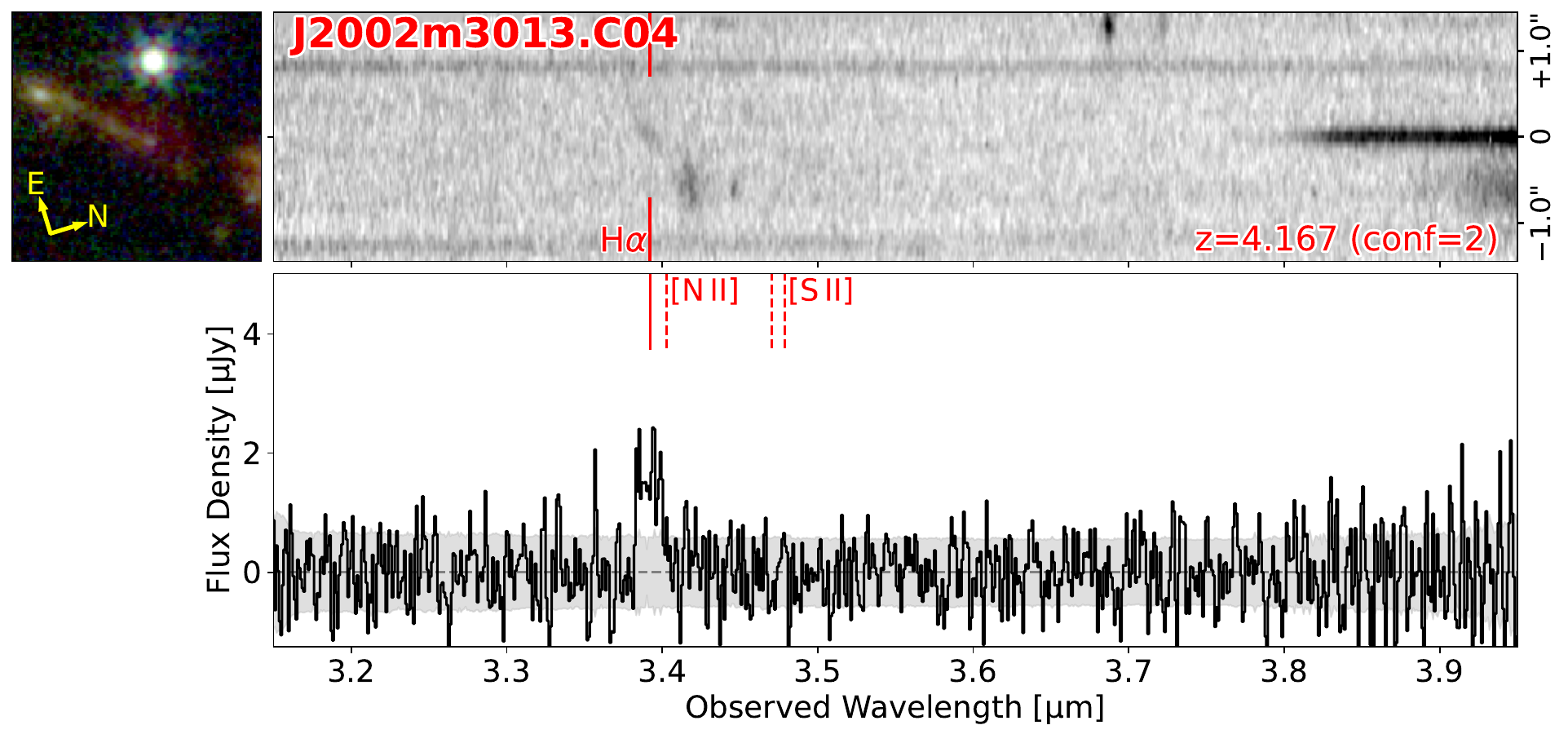}
\caption{Same as Figure~\ref{fig:spec_conf1} but for two DSFGs with spectroscopic redshifts at $4-6$ based on single emission line (\texttt{conf=2}).}
\label{fig:spec_conf2}
\end{figure*}

Figure~\ref{fig:spec_conf1} and Figure~\ref{fig:spec_conf2} show the 2D and 1D NIRCam grism spectra of these eight sources, grouped by the confidence level of the redshifts.

We classify the redshifts of six sources as secure (confidence \texttt{conf=1}; Figure~\ref{fig:spec_conf1}), as at least two spectral lines are detected (primary line at $\geq5\sigma$ and secondary line at $\geq3\sigma$).
Among them, the highest-redshift source J0109m3047.C02 is detected in \hb\ and \oiii\,$\lambda\lambda$4960, 5008, confirming the redshift at $z=5.549$.
We notice that J0109m3047.C02 is very close to a foreground galaxy ($z_\mathrm{phot}=1.18\pm0.05$, separation $\sim$\,0\farcs8).
We estimate a gravitational lensing magnification of $\mu = 1.91 \pm 0.50$ at the centroid of ALMA continuum emission (see details in Appendix~\ref{apd:01_lens}).
The other five sources in this group are detected in \nii\,$\lambda$6585 and/or \sii\,$\lambda\lambda$6718,6733 at $3\sigma$ significance at least, suggesting redshifts at $z = 4.07 - 4.68$.

The remaining two sources in Figure~\ref{fig:spec_conf2} only exhibit single-line detections in NIRCam grism spectra (confidence \texttt{conf=2}).
Without the detection of secondary line, it is hard to conclude their nature, but we argue that these emission lines are most likely \ha\ at $z=4.17$ and 4.80.
We measure the emission-line flux from the continuum-subtracted 1D spectra through Gaussian-profile fitting, and subtract the line fluxes from the F356W Kron-aperture photometry.
We then derive the equivalent widths (EWs) of these lines, finding large observed-frame EWs of 5240$\pm$500 and 890$\pm$270\,\AA\ for J1110m1329.C03 and J2002m3013.C04, respectively.
Such large EWs are extremely hard to be reproduced with lower-redshift Paschen or \hei\,$\lambda$10833 lines based on the analyses of F444W grism data of DSFGs in the GOODS-S field (FRESCO, \citealt{oesch23}; \citealt{boogaard23} and Sun, F., private communication), and the most likely line solution is therefore \ha\ or \oiii\,$\lambda$5008.
Therefore, we conclude that:

\textbullet\ J1110m1329.C03: The non-detection of a secondary line indicates that the detected line can only be \ha\ at $z = 4.798$, otherwise the \oiii\,$\lambda$4960 line would have been detected at $>3\sigma$.

\textbullet\ J2002m3013.C04: The wavelength of detected emission line of J2002m3013.C04 is very similar to the \ha\ wavelength of J2002m3013.C05 (velocity offset $\delta v$\,=\,350\,\si{km.s^{-1}}) in the same quasar field, and other line emitter at similar wavelengths are also identified in the vicinity of J2002m3013.C04 (see the 2D spectrum). Therefore, J2002m3013.C04 ($z=4.173$) is likely associated with a group of \ha\ emitters at $z=4.17$.

J1110m1329.C03 and J2002m3013.C04 are the only two single-line emitters in the ASPIRE DSFG sample that are undetected in the F115W band ($<3\sigma$) and thus likely reside at $z>4$, \textred{similar to many photometrically selected DSFGs at $z\gtrsim4$ through JWST imaging surveys \citep[e.g.,][]{mckinney23b,mckinney24,gottumukkala24,williams24}}. 
For all of the emission line solutions above, we also examine the ALMA spectral cubes to make sure that none of them are mis-identified \oiii$+$\cii\ emitters around quasar redshifts.
We also experiment other line solutions, e.g., \hei\,$\lambda$10833 and Pa$\gamma$, and we find that these solutions do not match the observed wavelengths or strengths of emission line complexes.

We notice that some of emission lines in Figure~\ref{fig:spec_conf1} and \ref{fig:spec_conf2} appear to be broad (FWHM\,$\gtrsim$\,1000\,\si{km.s^{-1}}). 
This is mostly because of morphological broadening with slitless spectroscopy, but not necessarily the presence of AGN. 
A galaxy with FWHM\,$\sim$\,0\farcs8 will exhibit a line with of $0\farcs8 \cdot (0\farcs063\,\si{pixel^{-1}})^{-1} (0.98\,\si{nm.pixel^{-1}}) \lambda_\mathrm{obs}^{-1} c \sim 1000$\,\si{km.s^{-1}} with NIRCam slitless spectroscopy, even without further consideration of instrumental or Doppler broadening. 
The presence (or not) of broad-line AGN will need to be examined with further slit or integral-field spectroscopy.

Table~\ref{tab:01_spec} summarizes the photometric and spectroscopic information of these eight DSFGs at $z = 4 - 6$.
We construct the JWST detection image by coadding F200W and F356W images, create image segments for regions that have at least 10 continuous pixels detected at $\geq5\sigma$, and then obtain aperture photometry using the Kron parameters \texttt{k=2.5} and \texttt{Rmin=1.2} in all JWST bands.
The photometry of J0109m3047.C02 is obtained on the residual images with other foreground galaxies modeled and subtracted with \textsc{galfit} (\citealt{galfit}; Appendix~\ref{apd:01_lens}).
The photometric uncertainty is estimated from sky background with a conservative noise floor of 5\% following the practice of NIRCam guaranteed-time-observation team (e.g., \citealt{tacchella23}).

At the ALMA wavelength, we adopt the peak flux densities of sources in the \textit{uv}-tapered images ($f_\mathrm{tap}$) multiplied by a factor of 1.1.
This is because we find that the scaled-up $f_\mathrm{tap}$ is systematically consistent with the flux densities measured with $r=1\arcsec$ aperture ($f_\mathrm{aper}$) and \textsc{casa} image-plane fitting (\texttt{imfit}; $f_\mathrm{imfit}$) in the native-resolution images, but with considerably smaller uncertainties.
The scaling factor is also verified with the injection experiment detailed in Appendix~\ref{apd:02_comp}.

For \ha\ emitters, we measure the emission-line fluxes through multi-component Gaussian-profile fitting.
For J0109m3047.C02 with complex morphological and thus line profile, we integrate the 1D spectrum within 20\,\AA\ (rest-frame) from the line center to derive the line fluxes.

\begin{figure*}[!t]
\centering
\includegraphics[width=\linewidth]{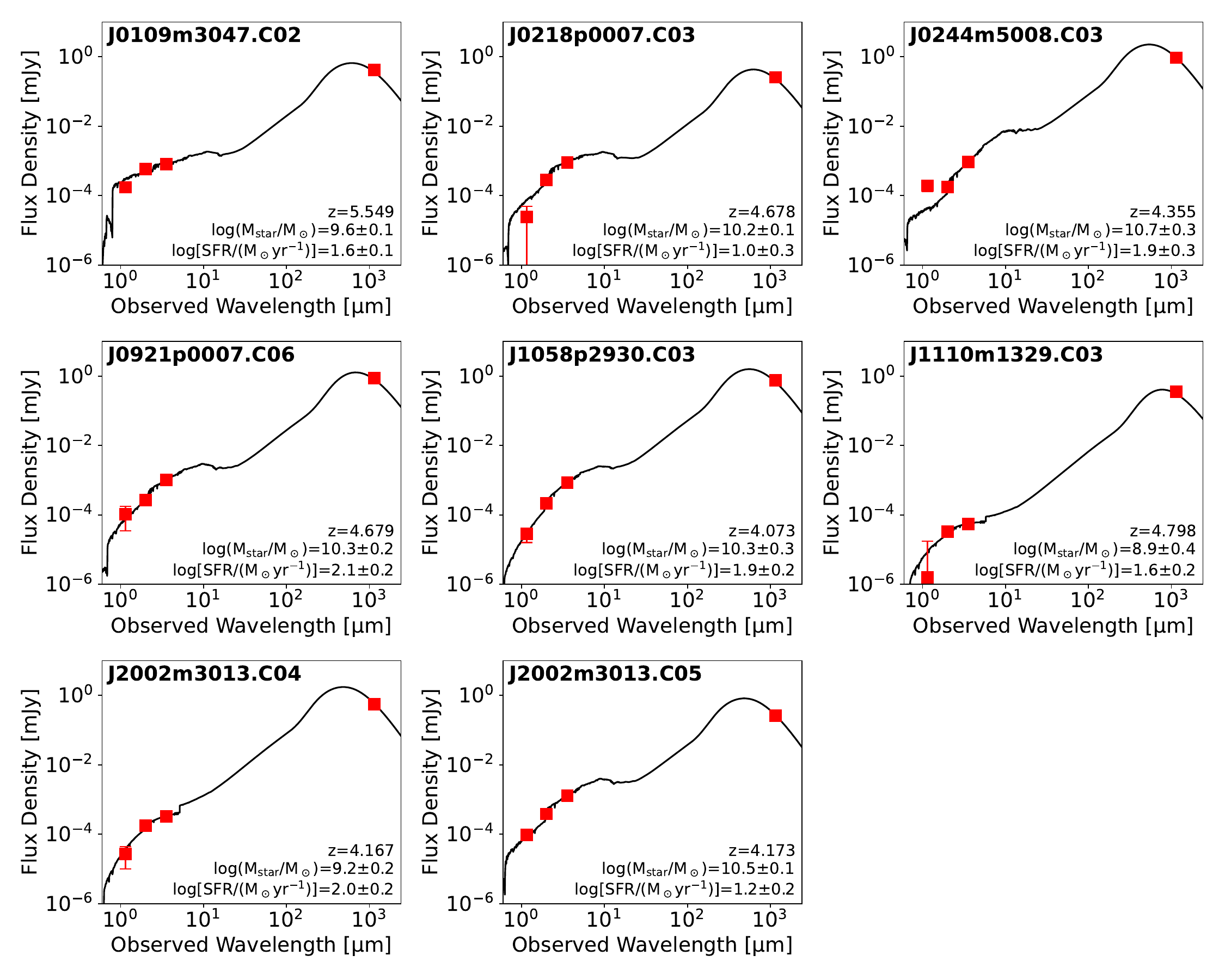}
\caption{JWST and ALMA SEDs of eight DSFGs in ASPIRE sample. Photometric measurements are shown as red squares. 
Best-fit SED models obtained with \textsc{CIGALE} are shown as solid black curves.
Source ID, redshifts and derived stellar masses, SFRs are indicated in the plots.
}
\label{fig:sed}
\end{figure*}

\begin{deluxetable*}{@{\extracolsep{0pt}}lrrrrcccrccc}[!th]
\tablecaption{Photometric and spectroscopic properties of ASPIRE DSFGs at $z=4-6$}
\label{tab:01_spec}
\tablewidth{0pt}
\tabletypesize{\footnotesize}
\tablehead{
\colhead{ID} & \colhead{R.A.} & 
\colhead{Decl.} & \colhead{$\mathrm{S/N}$} & \colhead{PB} & \colhead{$S_\mathrm{1.2mm}$} & \colhead{\zsp} & \colhead{conf} & \colhead{F115W} & \colhead{F200W} & \colhead{F356W} &  \colhead{$f_\mathrm{line}$} \\
\colhead{} & \colhead{[deg]} & \colhead{[deg]} & \colhead{} & \colhead{} & \colhead{[mJy]} & \colhead{}   & \colhead{} & \colhead{[mag]} & \colhead{[mag]} & \colhead{[mag]} & \colhead{[\si{10^{-18}.erg.s^{-1}.cm^{-2}}]} 
}
\startdata
J0109m3047.C02 & $ 17.46807$ & $-30.79361$ & 12.5 & 1.00 & 0.45$\pm$0.04 & 5.549 & 1 & 26.01$\pm$0.19 & 24.73$\pm$0.05 & 23.94$\pm$0.05 & 53.7$\pm$1.2 \\
J0218p0007.C03 & $ 34.70137$ & $  0.12652$ &  6.5 & 0.88 & 0.28$\pm$0.04 & 4.678 & 1 & $>26.70$       & 25.26$\pm$0.05 & 23.89$\pm$0.05 & 15.8$\pm$0.7 \\
J0244m5008.C03 & $ 41.01580$ & $-50.15628$ & 13.4 & 0.51 & 1.03$\pm$0.08 & 4.355 & 1 & 25.69$\pm$0.35 & 25.80$\pm$0.07 & 23.94$\pm$0.05 &  5.9$\pm$0.5 \\
J0921p0007.C06 & $140.33569$ & $  0.11316$ &  6.7 & 0.31 & 0.96$\pm$0.14 & 4.679 & 1 & $>25.60$       & 25.32$\pm$0.07 & 23.76$\pm$0.05 & 15.8$\pm$0.7 \\
J1058p2930.C03 & $164.54289$ & $ 29.50715$ &  6.6 & 0.31 & 0.84$\pm$0.13 & 4.074 & 1 & $>27.49$       & 25.58$\pm$0.05 & 24.03$\pm$0.05 &  3.6$\pm$0.7 \\
J1110m1329.C03 & $167.64528$ & $-13.49222$ &  9.1 & 0.99 & 0.38$\pm$0.04 & 4.798 & 2 & $>27.20$       & 27.62$\pm$0.22 & 26.42$\pm$0.05 &  6.0$\pm$0.6 \\
J2002m3013.C04 & $300.68137$ & $-30.22729$ & 11.2 & 0.76 & 0.61$\pm$0.05 & 4.167 & 2 & $>27.14$       & 25.77$\pm$0.07 & 24.99$\pm$0.05 &  7.6$\pm$2.3 \\
J2002m3013.C05 & $300.67392$ & $-30.22369$ &  7.0 & 1.00 & 0.29$\pm$0.04 & 4.173 & 1 & 26.44$\pm$0.14 & 24.94$\pm$0.05 & 23.54$\pm$0.05 & 13.2$\pm$0.9 \\
\enddata
\tablecomments{\texttt{S/N} is the maximum S/N measured from ALMA native-resolution and \textit{uv}-tapered map. \texttt{PB} is the ALMA primary beam response.
{$S_\mathrm{1.2mm}$} is the ALMA continuum flux density measured from the peak of \textit{uv}-tapered map, corrected for primary beam response and multiplied by a factor of 1.1 to match the flux densities measured from aperture and \texttt{CASA} image-plane fitting (\texttt{imfit}; Section~\ref{ss:03a_spec}).
The uncertainty for spectroscopic redshift is $\Delta z \sim 0.002$, mostly propagated from the wavelength calibration error of NIRCam WFSS.
$f_\mathrm{line}$ is the \oiii\,$\lambda$5008 (for J0109m3047.C02) or \ha\ (for other sources) line fluxes measured from 1D NIRCam grism spectra.
Lensing magnification is not corrected for J0109m3047.C02 (see Appendix~\ref{apd:01_lens}).
}
\end{deluxetable*}

\begin{deluxetable*}{@{\extracolsep{0pt}}lrrrrr}[!th]
\tablecaption{Physical properties of ASPIRE DSFGs at $z=4-6$ derived from SED modeling}
\label{tab:02_sed}
\tablewidth{0pt}
\tabletypesize{\footnotesize}
\tablehead{
\colhead{ID} & 
\colhead{$\log[M_\mathrm{star}/\mathrm{M}_{\odot}]$} & 
\colhead{$\log[\mathrm{SFR}_\mathrm{10Myrs}/(\mathrm{M}_{\odot}\,\mathrm{yr}^{-1})]$} &
\colhead{$\log[\mathrm{SFR}_\mathrm{UV}/(\mathrm{M}_{\odot}\,\mathrm{yr}^{-1})]$} &
\colhead{$\log[L_\mathrm{IR}/\mathrm{L}_{\odot}]$} & 
\colhead{$A_V$/mag} 
}
\startdata
J0109m3047.C02 &  9.6$\pm$0.1 & 1.6$\pm$0.1 & $ 0.9\pm$0.1 & 11.5$\pm$0.1 & 0.6$\pm$0.1 \\
J0218p0007.C03 & 10.2$\pm$0.1 & 1.0$\pm$0.3 &       $<0.4$ & 11.4$\pm$0.1 & 0.9$\pm$0.2 \\
J0244m5008.C03 & 10.7$\pm$0.3 & 1.9$\pm$0.3 & $ 0.8\pm$0.1 & 12.1$\pm$0.2 & 2.4$\pm$0.4 \\
J0921p0007.C06 & 10.3$\pm$0.2 & 2.1$\pm$0.2 &       $<0.9$ & 12.0$\pm$0.1 & 1.5$\pm$0.3 \\
J1058p2930.C03 & 10.3$\pm$0.3 & 1.9$\pm$0.2 & $-0.1\pm$0.2 & 11.9$\pm$0.1 & 2.0$\pm$0.3 \\
J1110m1329.C03 &  8.9$\pm$0.4 & 1.6$\pm$0.2 &       $<0.2$ & 11.4$\pm$0.2 & 2.1$\pm$0.5 \\
J2002m3013.C04 &  9.2$\pm$0.2 & 2.0$\pm$0.2 &       $<0.2$ & 11.8$\pm$0.2 & 1.5$\pm$0.2 \\
J2002m3013.C05 & 10.5$\pm$0.1 & 1.2$\pm$0.2 & $ 0.4\pm$0.1 & 11.6$\pm$0.1 & 1.4$\pm$0.3 \\
\enddata
\tablecomments{Light-weighted differential lensing magnification has been corrected for J0109m3047.C02 (see details in Appendix~\ref{apd:01_lens}).
$\mathrm{SFR}_\mathrm{10Myrs}$ is the SFR averaged over the last 10 Myrs of SFH, which is slightly different from the dust-obscured $\mathrm{SFR}_\mathrm{IR}$ (based on $L_\mathrm{IR}$) that we use to infer the obscured SFRD (Section~\ref{ss:04d_sfrd}).
}
\end{deluxetable*}

\subsection{SED Modeling}
\label{ss:03b_sed}

We obtain physical SED modeling of eight ASPIRE DSFGs at $z=4-6$ using software \textsc{CIGALE} \citep{cigale09,cigale19}.
We use four-band NIRCam and ALMA photometry for the fitting.
Because the rest-frame far-IR SEDs are only loosely constrained by ALMA photometry in one band, we have to rely on the energy balance assumption with \textsc{CIGALE} to infer the IR luminosity and thus obscured SFR.
In Section~\ref{ss:05a_fir}, we discuss the validity of energy balance assumption and impact from uncertain dust temperature.

We assume a commonly used delayed-$\tau$ star-formation history (SFH), in which $\mathrm{SFR}(t) \propto t \exp(-t / \tau)$ and $\tau$ is the peak time of star formation.
We allow the age of main stellar population to be 10--500\,Myr and $\tau$ to be 30--3000\,Myr. 
An optional late starburst is allowed in the last 10\,Myr of SFH, which could produce up to 20\% of total stellar mass.
We use \citet{bc03} stellar population synthesis models, and allow a metallicity range of 0.4$Z_{\odot}$--$Z_{\odot}$.
We adopt a modified \citealt{calzetti00} attenuation curve, and allow the variation of the power-law slope by $\pm$0.3 \textred{and $A_V$ effectively at 0--10}.
For simplicity, we assume \citet{casey12} mid-to-far-IR dust continuum model that could be parameterized by dust temperature ($T_\mathrm{dust}$; flat prior at 30--50\,K),
dust emissivity ($\beta_\mathrm{em}$, fixed at 1.8) and mid-IR power-law slope ($\alpha_\mathrm{MIR}$, fixed at 2.0).
Although the \citet{casey12} model does not include mid-IR polycyclic aromatic hydrocarbon (PAH) features, the inclusion of mid-IR power-law slope can well reproduce the luminosity excess from PAHs from a comparison of widely used empirical SED templates \citep[e.g.,][uncertainty $\lesssim$0.1\,dex]{chary01, rieke09}.
Note that we do not include nebular emission in the SED modeling for simplicity. 
Instead, we subtract the identified emission line fluxes (\ha, \oiii, \hb, \nii\ and \sii) from the F356W photometry, and perform the SED fitting for pure stellar and dust continuum.
For J0109m3047.C02, we correct for the differential lensing magnification in the SED modeling (Appendix~\ref{apd:01_lens}).

Figure~\ref{fig:sed} shows the best-fit SED models of eight DSFGs in our sample, and the derived physical parameters are presented in Table~\ref{tab:02_sed}.
We observe a median stellar mass $\log(M_\mathrm{star}/\si{M_\odot}) = 10.3 \pm 0.2$ for sources in our sample, and the median IR luminosity is $\log(L_\mathrm{IR} / \si{L_\odot}) = 11.7\pm0.1$, i.e., comparable to LIRGs in the local Universe.
We also derive UV SFR for each source through the observed F115W flux density (rest-frame $\sim$2000\,\AA) and the conversion $\mathrm{SFR}_\mathrm{UV} / (\si{M_\odot.yr^{-1}}) = 8.8\times10^{-29}\,L_\mathrm{FUV}/(\si{erg.s^{-1}.Hz^{-1}})$ (\citealt{md14} but assuming \citealt{chabrier03} IMF).
Under the same IMF assumption, the conversion between IR luminosity and dust-obscured SFR is $\mathrm{SFR}_\mathrm{IR} / (\si{M_\odot.yr^{-1}}) = 1.1\times10^{-10}\,L_\mathrm{IR}/\si{L_\odot}$, 
and the conversion between \ha\ luminosity and SFR is $\mathrm{SFR}_\mathrm{H\alpha} / (\si{M_\odot.yr^{-1}}) = 5\times10^{-42}\,L_\mathrm{H\alpha}/(\si{erg.s^{-1})}$ \citep{kennicutt98}.

\section{Results}
\label{sec:04_res}

\begin{figure*}[!t]
\centering
\includegraphics[width=\linewidth]{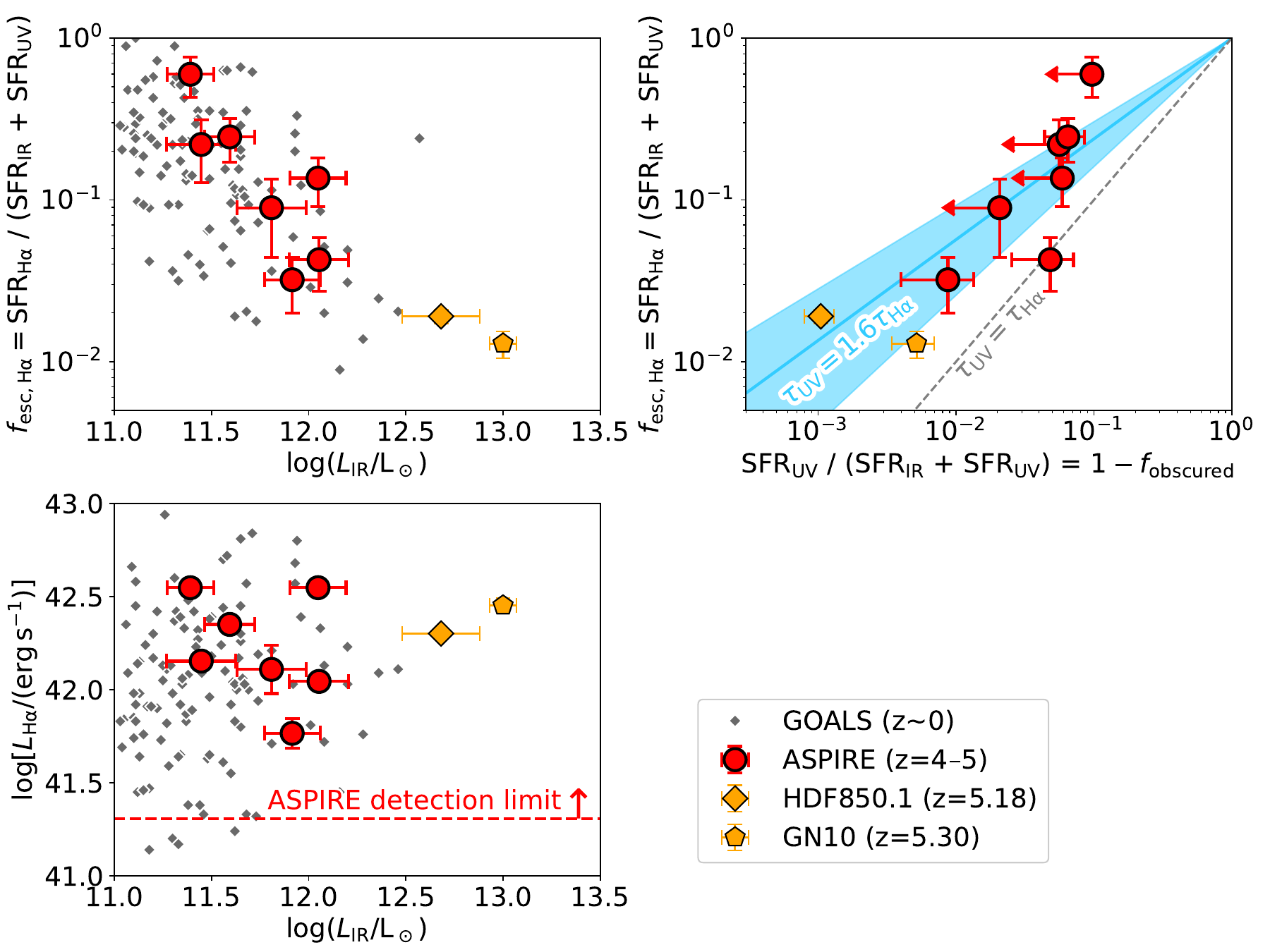}
\caption{The dust obscuration and escape of \ha\ photons from DSFGs. In the \textit{top-left} panel, we show the escape fraction of \ha\ photons versus IR luminosity. ASPIRE DSFGs are shown in red circles and two $z>5$ DSFGs in the FRESCO GOODS-N field (HDF850.1, GN10; measurements from \citealt{sun24}) are shown in orange symbols. Local (U)LIRGs in the GOALS sample are shown as small gray diamonds \citep{armus09,jinj19}.
DSFGs with higher \lir\ exhibit smaller escape fraction of \ha\ photons, resulting in a nearly constant \ha\ luminosity, i.e., $\log[L_\mathrm{H\alpha} / (\si{erg.s^{-1}})] \simeq 42.0 - 42.5$, across a $\sim$\,2-dex wide span of \lir\ as shown in the \textit{bottom-left} panel.
In the \textit{top-right} panel, we plot the escape fraction of \ha\ versus the escape fraction of UV ($1 - f_\mathrm{obscured}$).
The dashed gray line indicates the scenario that the optical depths at UV ($\tau_\mathrm{UV}$; $\sim$\,2000\,\AA) and \ha\ ($\tau_\mathrm{H\alpha}$) wavelength are identical.
The solid blue line and shaded region indicate the best-fit scaling relation and uncertainty $\tau_\mathrm{UV} = (1.61\pm0.34)\,\tau_\mathrm{H\alpha}$.
}
\label{fig:fesc_Ha}
\end{figure*}

\subsection{Obscured Fraction of UV and \ha\ Star Formation}
\label{ss:04a_fobs}

Most star formation in our ASPIRE DSFG sample at $z=4-6$ is obscured by dust.
If we define the obscured fraction of SFR as $f_\mathrm{obscured} = \mathrm{SFR}_\mathrm{IR} / (\mathrm{SFR}_\mathrm{IR} + \mathrm{SFR}_\mathrm{UV})$ \citep[e.g.,][]{whitaker17,fudamoto20}, then 96$\pm$2\% of their star formation is obscured by dust.
Such an obscured fraction of SFR is significantly higher than the ALPINE sample of \mstar-selected galaxies with comparable mass and redshifts ($f_\mathrm{obscured} = 67_{-7}^{+5}\%$; \citealt{fudamoto20}), suggesting that these ASPIRE DSFGs represent the dusty tail of the distribution of a wider star-forming galaxy population at this epoch.
The obscured fractions of SFR for ASPIRE DSFGs are still slightly less than that of the most extreme DSFG at $z\sim5$, e.g., HDF850.1 at $z=5.18$ \citep[]{hughes98, walter12}, whose obscured fraction is $f_\mathrm{obscured} \sim 99.9\%$ (\citealt{sun24}; see also \citealt{hd24}, \citealt{xiao24}).
\textred{We also notice that \citet{zimmerman24} suggests an increase of $f_\mathrm{obscured}$ at fixed stellar mass at higher redshifts in the \textsc{simba} cosmological simulations.}

Among the ASPIRE DSFG sample at $z=4-6$, five galaxies are undetected in the F115W band (rest-frame $\sim$\,2000\,\AA) and therefore they could be classified as HST-dark galaxies as introduced in Section~\ref{sec:01_intro}.
However, this is only a subset of 14 HST-dark galaxies selected with ALMA continuum detection and F115W non-detection ($<3\sigma$) among the full ASPIRE DSFG sample across all redshifts.
This suggests a complex and broad redshift distribution of HST-dark galaxies, and the photometric redshifts obtained in optical and near-IR can be strongly degenerate with the dust attenuation and stellar age of the galaxies \citep[e.g.,][]{huang11,caputi12,wangt16}.
Fortunately, the successful detections of \ha\ and \oiii\ lines of these highly obscured DSFGs through NIRCam WFSS provide the critical redshift confirmation, allowing us to study the dust attenuation in these DSFGs with much higher accuracy than sources without \zsp.

The top-left panel of Figure~\ref{fig:fesc_Ha} shows the escape fraction of \ha\ photons (defined as $f_\mathrm{esc,H\alpha} = \mathrm{SFR}_\mathrm{H\alpha} / (\mathrm{SFR}_\mathrm{UV} + \mathrm{SFR}_\mathrm{IR})$; not to confuse with the escape fraction in reionization/IGM context) versus \lir.
Assuming a constant SFH, $f_\mathrm{esc,H\alpha}$ is mainly controlled by the dust extinction 
(although caution that compared to IR luminosity, \ha\ luminosity is less sensitive to star formation at $\gtrsim$10\,Myr; \textred{see recent study of a $z=2.4$ post-starburst DSFGs by \citealt{cooper24}}).
It is clear that $f_\mathrm{esc,H\alpha}$ is strongly anti-correlated with \lir, both at $z\sim5$ or in the local Universe (as seen with the LIRGs/ULIRGs in the GOALS sample; \citealt{armus09}, \citealt{jinj19}).
Such an anti-correlation indicates that a larger fraction of \ha\ photons is dust-obscured in DSFG with more vigorous star formation activity. 
This is mainly driven by the fact that \ha\ luminosity of $z\sim5$ DSFG does not correlate with \lir\ (Spearman's $\rho=-0.05$, $p$-value\,=\,0.9) because of complex dust-stellar geometry \citep[e.g.,][]{gm09,ga22}.
As shown in the bottom-left panel of Figure~\ref{fig:fesc_Ha}, the \ha\ luminosity of $z\sim5$ DSFG remains almost constant (i.e., $\log[L_\mathrm{H\alpha} / (\si{erg.s^{-1}})] \simeq 42.0 - 42.5$) across an almost \,2-dex wide span of \lir.
Such a characteristic \ha\ luminosity range is well above the $5\sigma$ detection limit of ASPIRE (and also FRESCO; \citealt{oesch23}). 
Also as shown in the same plot, if local (U)LIRGs are placed at $z\sim5$, 96\% of them would be above our ASPIRE \ha\ luminosity detection threshold. Therefore, we conclude that the NIRCam grism spectroscopic surveys of high-redshift DSFGs are highly efficient and complete.

The escape fraction of UV photons (defined as $f_\mathrm{esc,UV} = 1 - f_\mathrm{obscured} = \mathrm{SFR}_\mathrm{UV} / (\mathrm{SFR}_\mathrm{IR} + \mathrm{SFR}_\mathrm{UV})$) is also controlled by dust extinction under the stable SFH assumption. 
\textred{A tight correlation between $f_\mathrm{esc,UV}$ and $f_\mathrm{esc,H\alpha}$ can demonstrate that these escape fractions are mostly driven by the dust obscuration instead of complex SFH because UV, \ha\ and far-IR emission is sensitive to SFH at different timescales.}
Therefore, we compare the escape fraction of \ha\ and UV photons in the top-right panel of Figure~\ref{fig:fesc_Ha}.
Both UV and \ha\ escape fractions are controlled by the optical depth at their wavelengths, i.e., $f_\mathrm{esc,UV} = \exp(-\tau_\mathrm{UV})$ and $f_\mathrm{esc,H\alpha} = \exp(-\tau_\mathrm{H\alpha})$.
We find that the typical $\tau_\mathrm{UV} / \tau_\mathrm{H\alpha}$ ratio of $z\sim5$ DSFGs is $1.61\pm0.34$.

Under \citet{calzetti00} extinction curve with negligible 2175-\AA\ bump, the dust optical depth ratio between UV ($\sim$\,2000\,\AA) and \ha\ wavelength is $\sim$\,2.7.
This is larger than the ratio derived above because nebular emission lines originate from stellar birth clouds with stronger dust attenuation, and in certain cases \ha\ from IR-luminous regions could be totally obscured \citep[e.g.,][]{ps15,colina23,bik24,sun24,uebler24}.
Therefore, the reduction factor to be applied on the nebular gas emission line $E(B-V)_g$ to derive stellar continuum $E(B-V)_s$ is $0.60\pm0.13$ for $z\sim5$ DSFGs, which is slightly higher than the canonical factor of 0.44$\pm$0.03 with \citet{calzetti00} extinction curve \citep{calzetti97}.
A larger reduction factor suggests that the difference of dust geometry between UV and nebular emission is smaller for $z\sim5$ DSFGs when compared to local starburst galaxies. 
This also increases the detectability of \ha\ emission of DSFGs at high redshifts. 
However, we also notice that the reduction factor is sensitive to the slope of underlying extinction curve. 
If the underlying extinction curve of high-redshift DSFGs is steeper, i.e., following the SMC curve \citep{gordon03}, then the reduction factor could be smaller (0.38$\pm$0.08).

\begin{figure*}
\centering
\includegraphics[width=\linewidth]{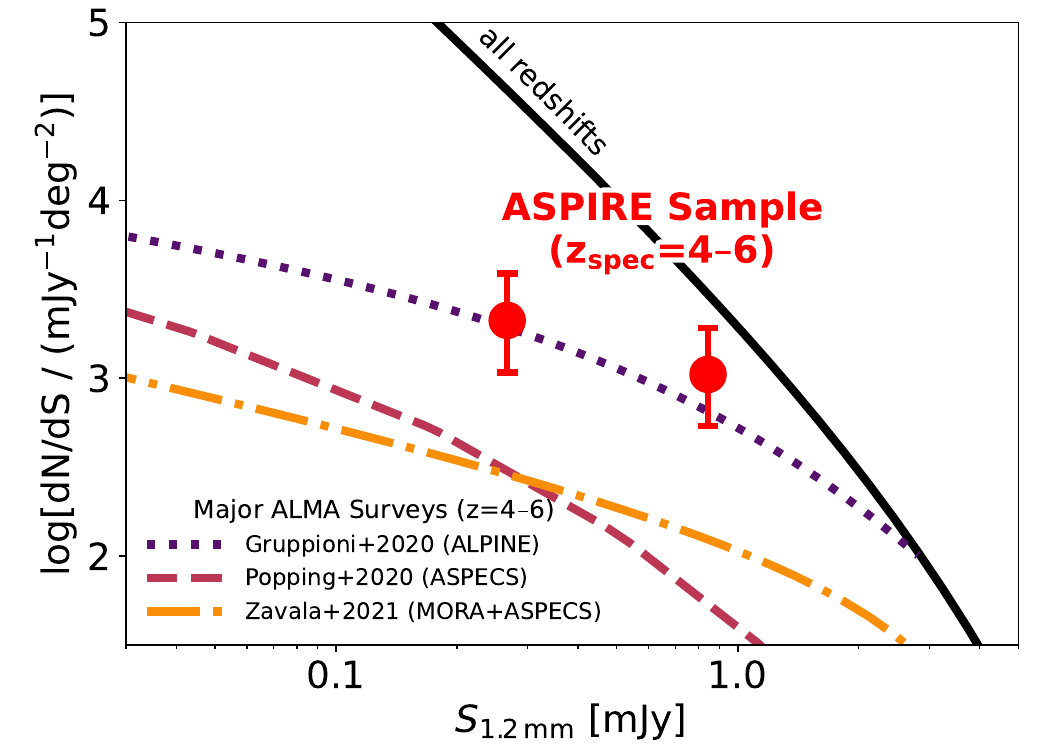}
\caption{Differential 1.2-mm number count of DSFGs at $z=4-6$ measured with the ASPIRE sample (solid red circles).
For comparison we show the 1.2-mm number count of DSFGs at all redshifts (\citealt{fujimoto24}; solid black line), and the number count of DSFGs at $z=4-6$ estimated based on \citet[][based on ALPINE survey; dotted purple line]{gruppioni20} and \citet[][based on MORA and ASPECS survey; dashed-dotted orange line]{zavala21} assuming their best-fit IRLFs and galaxy SEDs, and the semi–empirical model from \citet[][based on ASPECS survey; dashed maroon line]{popping20}.
}
\label{fig:count}
\end{figure*}

\subsection{Number Count of DSFGs at $z=4-6$}
\label{ss:04b_count}

One key focus of this study is to understand the number density of DSFGs at $z=4-6$ and their contribution to the IRLF and SFRD.
To measure these quantities, we first compute the effective survey area at above our detection limits in the native and tapered ALMA mosaics.
We make use of the RMS noises of these mosaics and the primary beam response maps to compute the effective survey area above each flux density limits, i.e., $A_\mathrm{max}(S_\mathrm{1.2\,mm})$.
Given the brightness of galaxies in our sample (observed $S_\mathrm{1.2mm}\geq 0.28$\,mJy; much larger than the $\sim$\,0.15\,mJy detection limit), these sources can be selected in most of our survey footprint ($A_\mathrm{max}$ ranges from 24 to 35\,arcmin$^2$ for each source).
Note that for the lensed source J0109m3047.C02, the $A_\mathrm{max}$ is computed using the lensing-corrected 1.2-mm flux density ($S_\mathrm{1.2mm} = 0.24$\,mJy).

By simply summing the $1/A_\mathrm{max}$ of ASPIRE DSFGs and correcting for the redshift desert at $z\simeq5.0-5.3$ where \ha\ or \oiii\ cannot be detected with NIRCam WFSS survey with the F356W filter, we find a surface density of $\sim1.1\times10^3$\,\si{deg^{-2}} for $z=4-6$ DSFGs at $S_\mathrm{1.2\,mm} \geq 0.24$\,mJy.
This is roughly twice of surface density of the HST $H$-dropout galaxies at $z \simeq 3 - 6$ and $S_\mathrm{0.87\,mm} \geq 0.6$\,mJy ($\sim$\,530\,\si{deg^{-2}}; \citealt{wangt19}).
The survey depth in terms of \lir\ is very similar for ASPIRE and the ALMA follow-up studies in \citet{wangt19}. 
Although the conventional $H$-dropout selection with HST and Spitzer/IRAC 
has identified obscured star formation that was previously missed by rest-frame UV surveys, such a selection
will also miss (\romannumeral1) $z>4$ DSFGs that are not totally obscured at HST wavelengths ($J$--$H$\,band), e.g., because of patchy dust geometry, and (\romannumeral2) highly dust-attenuated objects that are faint at 3--5\,\micron\ and below the Spitzer/IRAC detection limit.

We measure the differential number count ($dN/dS$) of ASPIRE DSFGs at $z=4-6$ through the $1/A_\mathrm{max}$ method:
\begin{equation}
\frac{dN}{dS} = \frac{1}{d\log S} \sum_i \frac{1}{C_i A_\mathrm{max}(S_i) }
\end{equation}
where $d\log S$ is the flux density bin size in log scale, $C_i$ is the completeness for the $i$-th source in the bin, and $A_\mathrm{max}(S_i)$ is the effective survey area at above the source flux density $S_i$.

We present the completeness correction simulation in Appendix~\ref{apd:02_comp}, and as a quick summary, our ALMA continuum source selection is highly complete ($>$\,95\% for all sources) because of high fidelity of detections ($\mathrm{S/N} \geq 6.5$).
Strictly speaking, because of multiple redshift gaps with NIRCam WFSS survey in the F356W band (Section~\ref{ss:03a_spec}), the full ASPIRE DSFG sample is not spectroscopically complete across all redshifts.
However, we argue that we have reached a high spectroscopic completeness at $z=4-5$ and 5.3--6 because the \ha\ and \oiii\ luminosity of sources in our sample are well above the $5\sigma$ detection limit (indicated by the dashed red line in low-left panel of Figure~\ref{fig:fesc_Ha}).
Even if the completeness of \oiii\ spectroscopy for DSFGs cannot be properly assessed because of limited JWST data taken so far (high-redshift DSFGs may exhibit a diverse distribution of \oiii\ strengths; e.g., \citealt{mcKinney23,barrufet24,sun24}), we argue that the number density of DSFGs at $z>5$ should be naturally smaller than that at $z<5$, same as we observed.
Therefore, the true number density of DSFGs at $z=4-6$ should be close to our measurement, or just slightly higher.

We set two flux density bins at 0.15--0.47 and 0.47--1.5\,mJy (bin size is 0.5\,dex).
We also run Monte Carlo (MC) simulation to draw 1.2-mm flux density from the observed values and errors, compute the corresponding $A_\mathrm{max}$ and completeness and then derive the number count and its uncertainty.
The uncertainty includes Poisson error following the prescription of \citet{gehrels86}, and we also include the error from cosmic variance (shot noise) following the prescription of \textred{\citet[][same as the method adopted by \citealt{mckinney24} for DSFGs in the COSMOS-web field]{moster11}}, which is computed using the geometry of survey footprint \textred{(25 fields with area of 1\farcm2$\times$1\farcm2)}, redshift interval \textred{($\Delta z = 1.7$ around $z=5$)}, dark matter correlation function, $\sigma_8$ and galaxy bias at $z\sim5$ \textred{(0.17 and and $\sim$8)}.
The uncertainty from cosmic variance ($\sim$0.05\,dex) is much smaller than that from Poisson error ($\sim$0.28\,dex in each bin), and we also experiment other cosmic variance prescriptions and find similar results \citep[0.05--0.07\,dex;][]{trenti08,driver10}.

Figure~\ref{fig:count} shows the 1.2-mm number of DSFGs at $z=4-6$ measured from the ASPIRE survey (red circles).
We highlight the fact that the only noticeable uncertainty of our measurement is from Poisson statistics, nor cosmic variance or any other astrophysical assumption (e.g., \zph\ or SED model).
The number count of $z=4-6$ DSFGs seems to be much flatter than the 1.2-mm number count of DSFGs across all redshifts (solid black line, \citealt{fujimoto24}; see also \citealt{fujimoto16,gl20,gomez22,chenj23,adscheild24} that cover similar $S_\mathrm{1.2mm}$ space).
This indicates a larger fraction of DSFGs at higher redshifts toward the bright end of millimeter number count function \citep[e.g.,][]{bethermin17, casey18, casey21, lagos20, sun22a, chenc22}.

For comparison, we also show the number count of DSFGs at $z=4-6$ based on the semi-empirical model from \citet[][based on ASPECS survey]{popping20} and IRLFs modeled by \citet[][based on ALPINE survey]{gruppioni20} and \citet[][based on MORA and ASPECS survey]{zavala21}.
To convert IRLFs in literature to number counts, we assume similar far-IR SEDs as those assumed in these original works.
In \citet{zavala21}, the authors derived a best-fit dust emissivity $\beta_\mathrm{em}\sim 1.8$ from their backward modeling, and they assumed a relation between the rest-frame wavelength of far-IR SED peak and IR luminosity ($\lambda_\mathrm{peak} - L_\mathrm{IR}$; \citealt{casey18}), which corresponds to a modified blackbody (MBB) $T_\mathrm{dust} \sim 32$\,K at $L_\mathrm{IR} = 10^{12}$\,\lsun\ through \citealt{casey12} parametrization.
In \citet{gruppioni20}, the authors made use of best-fit far-IR SED through multiple templates included in the software \textsc{Le\,Phare} \citep{lephare}.
Therefore, we convert their best-fit IRLF at $z=3.5-6$ to 1.2-mm number count assuming the \citet{chary01} far-IR SED templates, one of the major template sets adopted by \textsc{Le Phare} (effectively $T_\mathrm{dust} \sim 45$\,K and $\beta_\mathrm{em} \sim 1.5$ for $S_\mathrm{1.2mm} \sim 0.5$\,mJy sources at $z\sim 5$).

The number count of $z=4-6$ DSFGs from ASPIRE sample is similar to that of \citet{gruppioni20} based on non-target ALMA Band-7 continuum galaxies discovered through ALPINE survey \citep{bethermin20, lefevre20}.
Because the primary targets of ALPINE survey are star-forming galaxies at $z \sim 5$, it has been speculated that the excess of $z=4-6$ DSFG number count from ALPINE (relative to ASPECS) may arise from the clustering of non-target DSFGs around their main targets \citep[e.g.,][]{zavala21}.
Our observations suggest that the number density of $z=4-6$ DSFG measured from ALPINE survey is sufficiently accurate despite uncertainties with \zph.

Our measurements are higher than the models based on ASPECS survey \citep{popping20, zavala21} by $\sim$\,1\,dex.
Such a large discrepancy cannot be explained by any form of error from the ASPIRE sample.
We think the large difference is caused by the strong shot noise with the ASPECS sample ($\sigma \sim 0.5$\,dex, including both Poisson error and cosmic variance).
\citet{aravena20} showed that no DSFG at $z>4$ was blindly discovered with the ASPECS deep 1-mm or 3-mm survey.
In contrast, the eight DSFGs in our sample are discovered along 7 out of 25 quasar sightlines, and therefore the chance to detect one DSFG at $z=4-6$ within the 1.4-arcmin$^2$ survey area per sightline is only 28\%.
The survey area of ASPECS in the HUDF is $\sim$\,3 times of a typical ASPIRE field/sightline, and thus the chance to detect zero DSFG at $z=4-6$ is $(1-28\%)^3\sim37$\%, a non-negligible probability.
Therefore, we attribute the discrepancy between ASPIRE and ASPECS measurements to the underdensity of $z=4-6$ DSFG in the HUDF region.
Similar conclusions have been drawn from previous studies for DSFGs \citep[e.g.,][]{fujimoto24} and even for Lyman-break galaxies and AGN \citep[e.g.,][]{cowie02, moretti03, bauer04, oesch07}.

\begin{figure*}[!t]
\centering
\includegraphics[width=\linewidth]{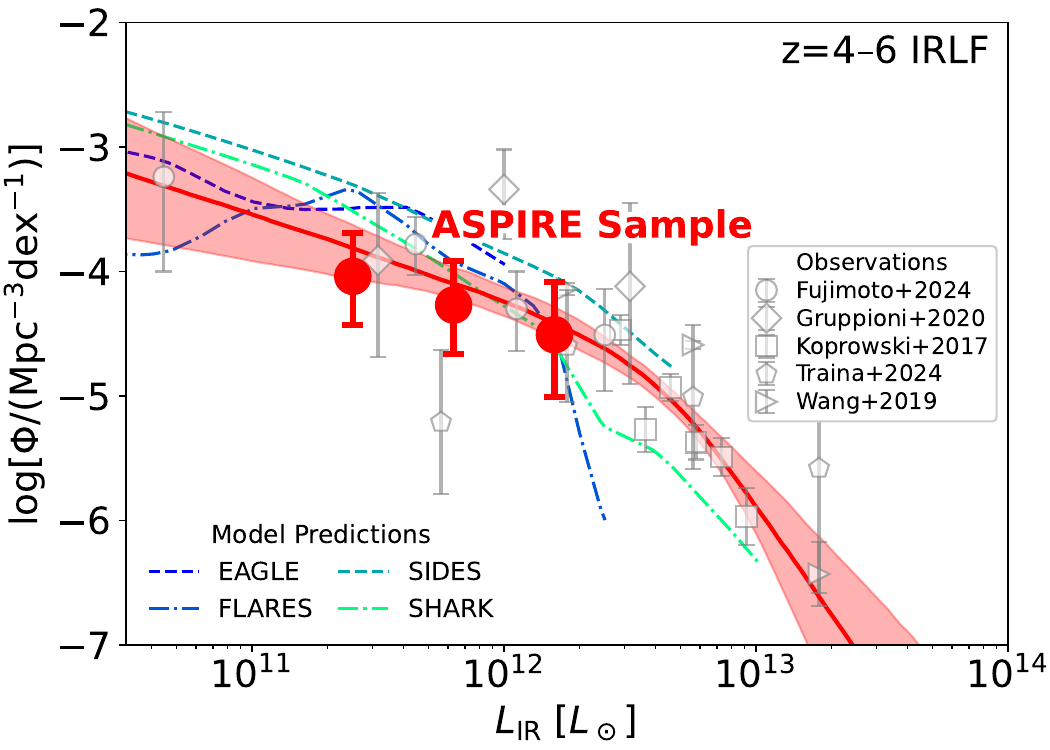}
\caption{Spectroscopically complete infrared luminosity function at $z=4-6$ measured with the ASPIRE sample (solid red circles).
The best-fit IRLF and its uncertainties (16--84th percentile) are indicated with the solid red line and shaded region, respectively.
For comparison we show the direct measurements from previous observations (incomplete with spectroscopy) as open gray symbols, which include \citet{koprowski17}, \citet{wangl19}, \citet{gruppioni20}, \citet{fujimoto24}, \citet{traina24}.
Theoretical model predictions are shown as colored lines including EAGLE \citep{trayford20}, FLARES \citep{vijayan22}, SIDES \citep{bethermin22} and SHARK \citep{lagos20}.
}
\label{fig:IRLF}
\end{figure*}

\begin{table*}[!t]
\caption{Infrared luminosity function at $z=4-6$ measured with ASPIRE DSFGs \label{tab:03_irlf}}
\centering
\begin{tabular}{@{\extracolsep{4pt}}rrrr}
\hline\hline
\multicolumn{4}{l}{Observed infrared luminosity function} \\\hline
$\log(L_\mathrm{IR}/\si{L_\odot})$ & 11.2--11.6 & 11.6--12.0 & 12.0--12.4 \\
$<N>$                              &       2.71 &       2.88 &       1.63 \\
$\log[\Phi / (\mathrm{Mpc}^{-3}\mathrm{dex}^{-1})]$  &  $-4.04_{-0.39}^{+0.35}$ &  $-4.27_{-0.39}^{+0.35}$ &  $-4.51_{-0.50}^{+0.42}$ \\
\hline\hline
\multicolumn{4}{l}{Best-fit double-power-law parameters} \\\hline
$\log[\Phi^{\star} / (\mathrm{Mpc}^{-3}\mathrm{dex}^{-1})]$ & $\log(L_\mathrm{IR}^{\star} / L_{\odot})$ & $\alpha$ & $\beta$ \\
$-4.52_{-0.55}^{+0.49}$ & $12.58_{-0.29}^{+0.26}$ & $0.59_{-0.45}^{+0.39}$ & $3.24_{-0.95}^{+1.63}$ \\
\hline\hline
\end{tabular}
\tablecomments{\small $\log(L_\mathrm{IR}/\si{L_\odot})$ is the luminosity bin, $<N>$ is the average number of sources in the luminosity bin through MC simulations, and $\log[\Phi / (\mathrm{Mpc}^{-3}\mathrm{dex}^{-1})]$ is the measured volume density of sources in the luminosity bin.
Parameters for the IRLF are derived assuming a double-power-law function through \textsc{emcee} (Section~\ref{ss:04c_irlf}).
}
\end{table*}

\begin{figure*}[!t]
\centering
\includegraphics[width=\linewidth]{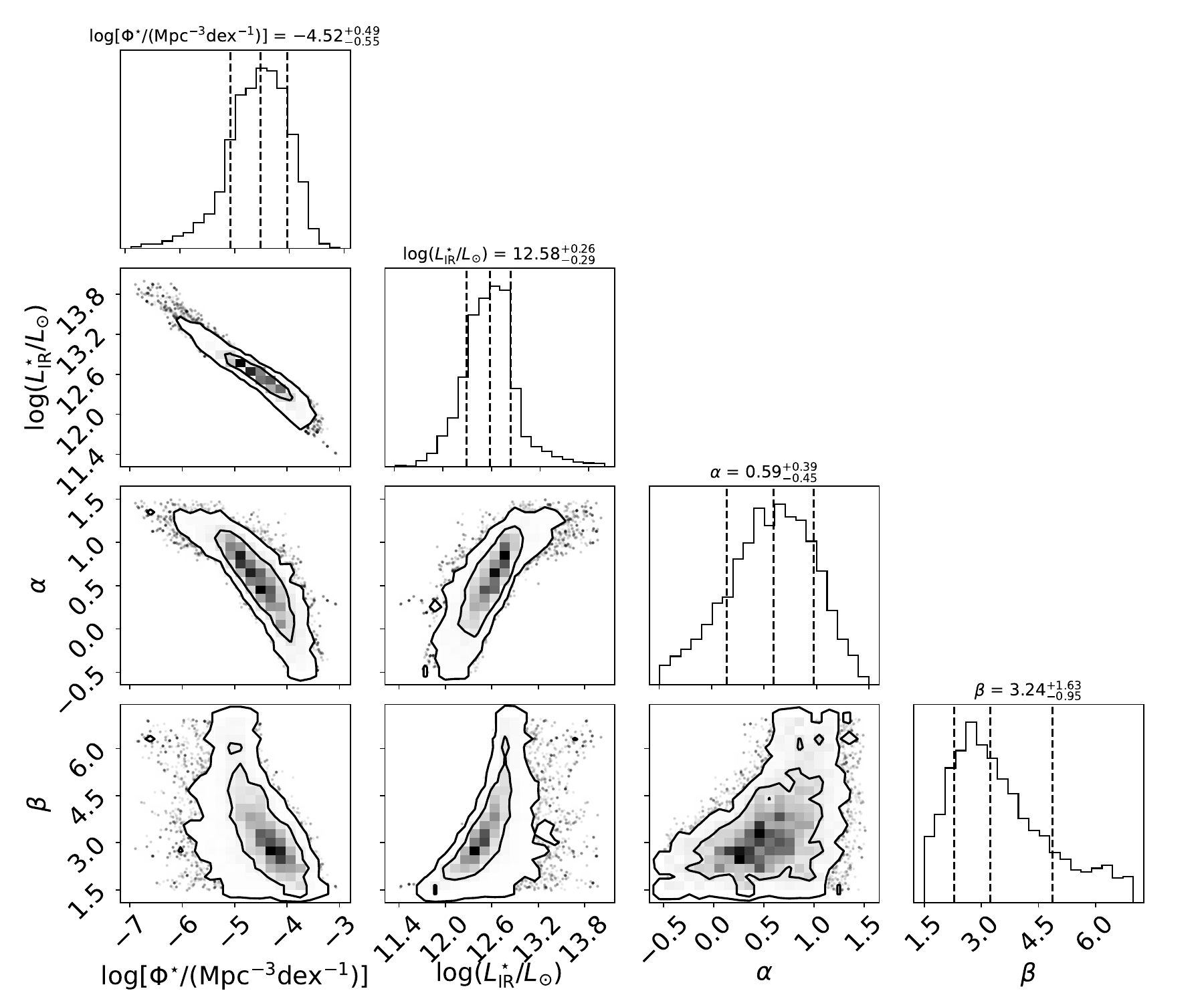}
\caption{MCMC corner plot of the double-power-law parameters for the IRLF fitting (Figure~\ref{fig:IRLF} and Section~\ref{ss:04c_irlf}). Contours are at $1-2\sigma$.
}
\label{fig:corner}
\end{figure*}

\subsection{Spectroscopically Complete Infrared Luminosity Function}
\label{ss:04c_irlf}

We further measure the IRLF at $z=4-6$ from the ASPIRE DSFG sample.
Such an IRLF is spectroscopically complete down to $L_\mathrm{IR}\sim2\times10^{11}$\,\lsun\ and almost free from the impact of cosmic variance.
Similar to the method present in Section~\ref{ss:04b_count}, we adopt the $1/V_\mathrm{max}$ method \citep{schmidt68}:
\begin{equation}
\Phi(L) = \frac{1}{d\log L} \sum_i \frac{1}{C_i V_\mathrm{max}(L_i) }
\end{equation}
where $d\log L$ is the luminosity bin size in log scale, $C_i$ is the completeness for the $i$-th source in the bin, and $V_\mathrm{max}(L_i)$ is the effective survey volume at above the source luminosity $L_i$.
We make use of $L_\mathrm{IR}$ and uncertainties measured from \textsc{CIGALE} energy-balance SED fitting.
The uncertainty of \lir\ and IRLF propagated from loosely constrained $T_\mathrm{dust}$ is further discussed in Section~\ref{ss:05a_fir}.

We set three \lir\ bins at $10^{11.2} - 10^{11.6}$, $10^{11.6} - 10^{12.0}$ and $10^{12.0} - 10^{12.4}$\,\lsun. 
Similarly to Section~\ref{ss:04b_count}, we also run MC simulation to randomly draw \lir\ from the SED-derived \lir\ and uncertainties.
The ratio of drawn \lir\ and best-fit \lir\ is applied to scale the ALMA S/N for the completeness and effective survey volume calculation.
The uncertainty of our luminosity function measurements include the standard deviation from MC simulation, Poisson error (following \citealt{gehrels86}) and cosmic variance ($\sim$\,0.03\,dex, negligible; following \citealt{moster11} prescription), respectively.
The derived IRLF is presented in Table~\ref{tab:03_irlf} and Figure~\ref{fig:IRLF}.

At the \lir\ range probed by ASPIRE DSFGs, 
the availability of measurements in literature is limited. 
This is because these DSFGs are typically below the detection limit of single-dish telescope at (sub)-millimeter wavelengths (e.g., JCMT/SCUBA2 and Herschel/SPIRE).
Our measurements are consistent with those of \citet{fujimoto24} based on ALCS data in lensing cluster fields, and also broadly consistent with the measurements from \citet{gruppioni20} and \citet{traina24} in their lowest-luminosity bins.
Nevertheless, we observe a flattening of IRLF toward the faint end, and the volume density of LIRGs at $z=4-6$ is somewhat lower than many model predictions \citep[e.g.,][]{lagos20,trayford20,bethermin22,vijayan22}.

We fit the measured IRLF with a double-power-law model following \citet{fujimoto24}:
\begin{equation}
\Phi(L) = \Phi^{\star} \Big[\Big(\frac{L}{L^\star}\Big)^\alpha + \Big(\frac{L}{L^\star}\Big)^\beta \Big]^{-1}
\end{equation}
where $L^{\star}$ and $\Phi^{\star}$ are the characteristic luminosity and volume density, respectively. 
We define $\alpha$ and $\beta$ as the faint-end and bright-end slope of LF, respectively.
We obtain Monte Carlo Markov Chain (MCMC) fitting of the function above using \textsc{emcee} \citep{emcee}.
Because ASPIRE data do not constrain the bright end of IRLF toward $\sim10^{13}$\,\lsun, we have to include literature measurements from \citet{koprowski17}, \citet{gruppioni20}, \citet{fujimoto24} and \citet{traina24} into our fitting.
All literature samples included here are not spectroscopically complete. 
Therefore, we artificially increase their uncertainties by a factor of $\sqrt{2}$ to reduce their weighting in the fitting and compensate for potentially underestimated error from \zph.
The allowed ranges for the double-power-law parameters are $-10 < \log(\Phi^\star) < 0$, $11 < \log(L^\star) < 14$, $-0.5 < \alpha < 1.5$ and $1.5 < \beta < 7.0$.
All prior distributions are flat.

The best-fit IRLF and its uncertainties (16--84th percentile) are indicated with the solid red line and shaded region in Figure~\ref{fig:IRLF}, respectively.
The best-fit parameters of IRLF and uncertainties are also presented in Table~\ref{tab:03_irlf}.
We find that the combined ASPIRE and ALCS measurements yield a flattened faint-end slope of $\alpha = 0.59_{-0.45}^{+0.39}$, in between the faint-end slope measured by \citet[][$\sim0.4$]{zavala21} and \citet[][$\sim0.9$]{fujimoto24}.
Although large uncertainties are seen with all of these parameters, as shown in the corner plot of our MCMC fitting (Figure~\ref{fig:corner}), $\log(\Phi^\star)$, $\alpha$ and $\beta$ are all strongly degenerate with $\log(L^\star)$.
Therefore, the integral of the IRLF (i.e., obscured SFRD) can be measured relatively accurately (Section~\ref{ss:04d_sfrd}).
Future spectroscopic constraints of $z=4-6$ IRLF at $L_\mathrm{IR} \gtrsim L^\star$ will be very helpful to accurately determine $L^\star$ and other parameters (Section~\ref{ss:05b_bend}).

\begin{figure*}[!t]
\centering
\includegraphics[width=\linewidth]{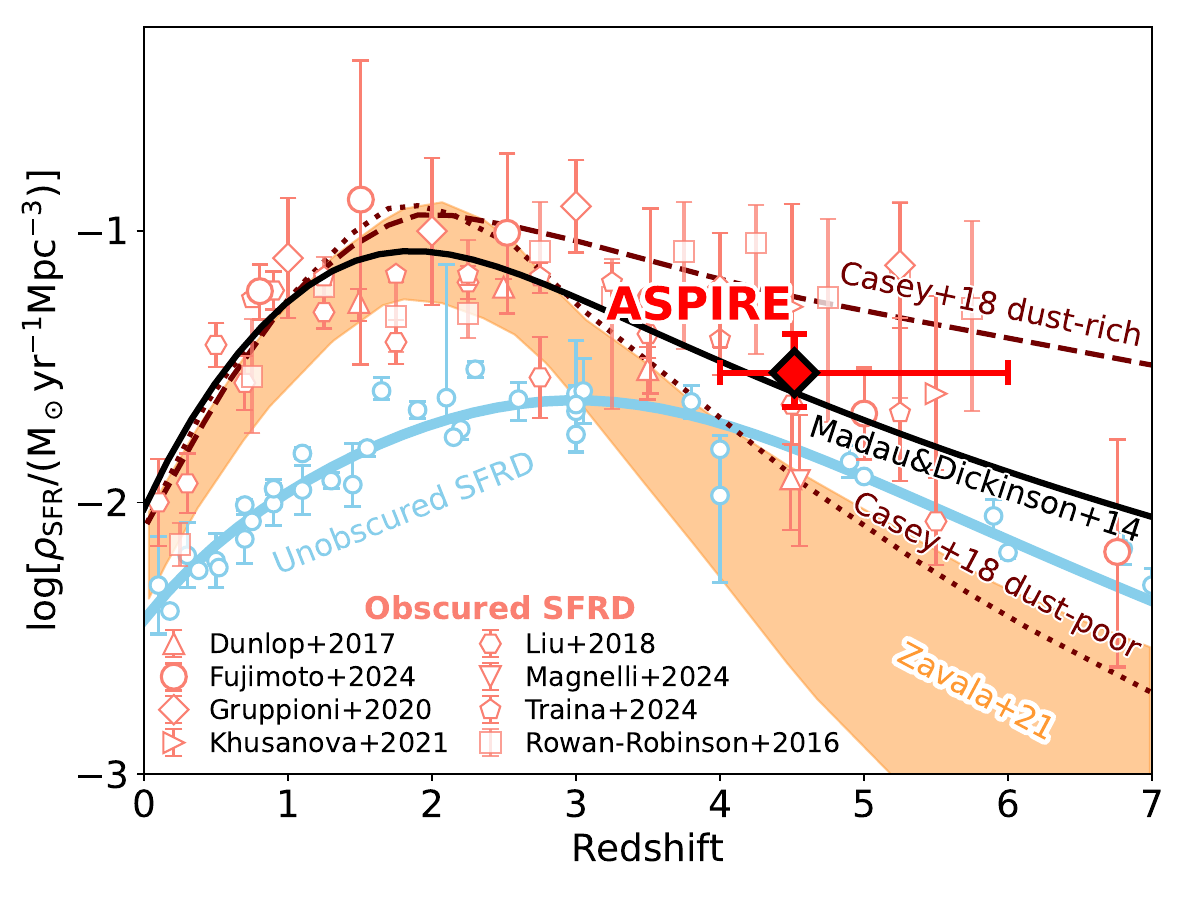}
\caption{Obscured cosmic star formation rate density measured with the ASPIRE sample at $z=4-6$ (red diamond).
For comparison we plot literature measurements of obscured SFRD as open salmon-pink symbols, including \citet{rr16}, \citet{dunlop17}, \citet{liu18}, \citet{gruppioni20}, \citet{khusanova21}, \citet{fujimoto24}, \citet{magnelli24} and \citet{traina24}.
Unobscured SFRD measured in rest-frame UV are shown as open sky-blue circles (including \citealt{reddy09}, \citealt{cucciati12}, \citealt{bouwens15}, \citealt{bouwens20}, \citealt{finkelstein15}, \citealt{alavi16}, \citealt{mehta17}, \citealt{moutard20},   \citealt{sunlei23}) and the best-fit redshift-evolution model as indicated by the solid sky-blue curve.
The \citet{casey18} ``dust-rich'' and ``dust-poor'' scenarios are shown as dashed and dotted maroon lines, respectively.
The inferred obscured SFRDs from a backward evolution model of 1--3\,mm number counts \citep{zavala21} are shown as orange shaded region.
The solid black line shows the best-fit total SFRD from \citet{md14}.
\citet{chabrier03} IMF has been assumed for all SFRD measurements in this plot.
}
\label{fig:sfrd}
\end{figure*}

\subsection{Obscured Star-Formation Rate Density}
\label{ss:04d_sfrd}

We compute the obscured cosmic SFRD using the derived IRLF. 
Similar to \citet{fujimoto24}, we integrate the IRLF down to $L_\mathrm{IR} = 10^{10}$\,\lsun, which corresponds to $\sim0.0025 L^\star$.
Because of the flattening of IRLF toward the faint end, the integral of IRLF down to $L_\mathrm{IR} = 10^{9}$\,\lsun\ is only higher than that down to $L_\mathrm{IR} = 10^{10}$\,\lsun\ by $\sim0.03$\,dex.
We convert the integral of IRLF to obscured SFRD using the factor listed in Section~\ref{ss:03b_sed}.
We derive an obscured SFRD of $\log[\rho_\mathrm{SFR,IR}/(\si{M_\odot.yr^{-1}.Mpc^{-3}})] = -1.52_{-0.13}^{+0.14}$ at $z=4-6$ (median $z = 4.5$).

Figure~\ref{fig:sfrd} shows the obscured and unobscured SFRD measurements at $z\simeq0 - 7$ in the literature, and the ASPIRE measurement is highlighted as the red diamond.
At $z\gtrsim4$, previous determinations of obscured SFRD are highly uncertain with a wide dispersion over 1-dex span \citep[e.g.,][]{rr16,dunlop17,koprowski17,liu18,gruppioni20,khusanova21,zavala21,magnelli24,traina24}.
Our determination is within the dispersion of these previous measurements, and specifically, our measurement is about five times of the obscured SFRD derived by \citet{zavala21} through a backward modeling of millimeter number counts.
As mentioned in Section~\ref{ss:04b_count}, we caution the strong cosmic variance and void of $z>4$ DSFGs in the HUDF region that could affect the number counts at 1 and 3\,mm used by \citet{zavala21}.
The uncertainty from far-IR SED can also lead to the error of SFRD determination with both the backward modeling method \citep{casey18, zavala21} and also our direct determination, which will be further discussed in Section~\ref{ss:05a_fir}.

We also compare our measurement with the so-called ``dust-poor'' and ``dust-rich'' scenarios of obscured SFRD evolution proposed by \citet{casey18}.
Our result is in between these two model predictions at $z\sim4.5$.
The ASPIRE obscured SFRD is also comparable to the total SFRD (UV\,$+$\,IR) at $z\sim4.5$ from \citet{md14} best-fit model ($\log[\rho_\mathrm{SFR,total}/(\si{M_\odot.yr^{-1}.Mpc^{-3}})] = -1.60$).
Similar conclusion has also been drawn by recent ALCS observations \citep{fujimoto24}, suggesting that the total SFRD at $z=4-6$ could have been underestimated previously.

To understand the total amount and obscured fraction of SFRD at $z=4-6$, we compile the unobscured SFRD measurements from rest-frame UV surveys at $z=0-7$ \citep[][]{reddy09, cucciati12, bouwens15, bouwens20, finkelstein15, alavi16, mehta17, moutard20, sunlei23}.
Following \citet{md14}, we fit the redshift evolution of unobscured SFRD with the following function:
\begin{equation}
\rho_\mathrm{SFR,UV}(z) = \Psi_0 \frac{(\frac{1+z}{1 + z_0})^a}{[1 + (\frac{1+z}{1 + z_0})^b]}\,\si{M_\odot.yr^{-1}.Mpc^{-3}}
\end{equation}
and we derive $z_0 = 3.74\pm0.27$, $\log(\Psi_0)=-1.37\pm0.03$, $a=1.57\pm0.16$ and $b = 5.85 \pm 0.33$.
This implies an unobscured SFRD of $\log[\rho_\mathrm{SFR,UV}/(\si{M_\odot.yr^{-1}.Mpc^{-3}})] = -1.80$ at $z\sim4.5$, about half of the obscured SFRD as we derived based on ASPIRE data.
Therefore, we conclude that $66_{-7}^{+7}$\%\ of the total SFRD is obscured by dust at this epoch.
The total SFRD that we derive is $\log[\rho_\mathrm{SFR,total}/(\si{M_\odot.yr^{-1}.Mpc^{-3}})] = -1.34_{-0.08}^{+0.10}$, which is $180_{-30}^{+46}$\%\ of the total SFRD from \citet{md14} best-fit model (similar to the conclusion in \citealt{fujimoto24}).

\section{Discussion}
\label{sec:05_dis}

In this section, we discuss the potential caveats of our determination of IRLF and SFRD. 
We also propose improvements that could be made through observations in the near future.

\begin{figure}[!t]
\centering
\includegraphics[width=\linewidth]{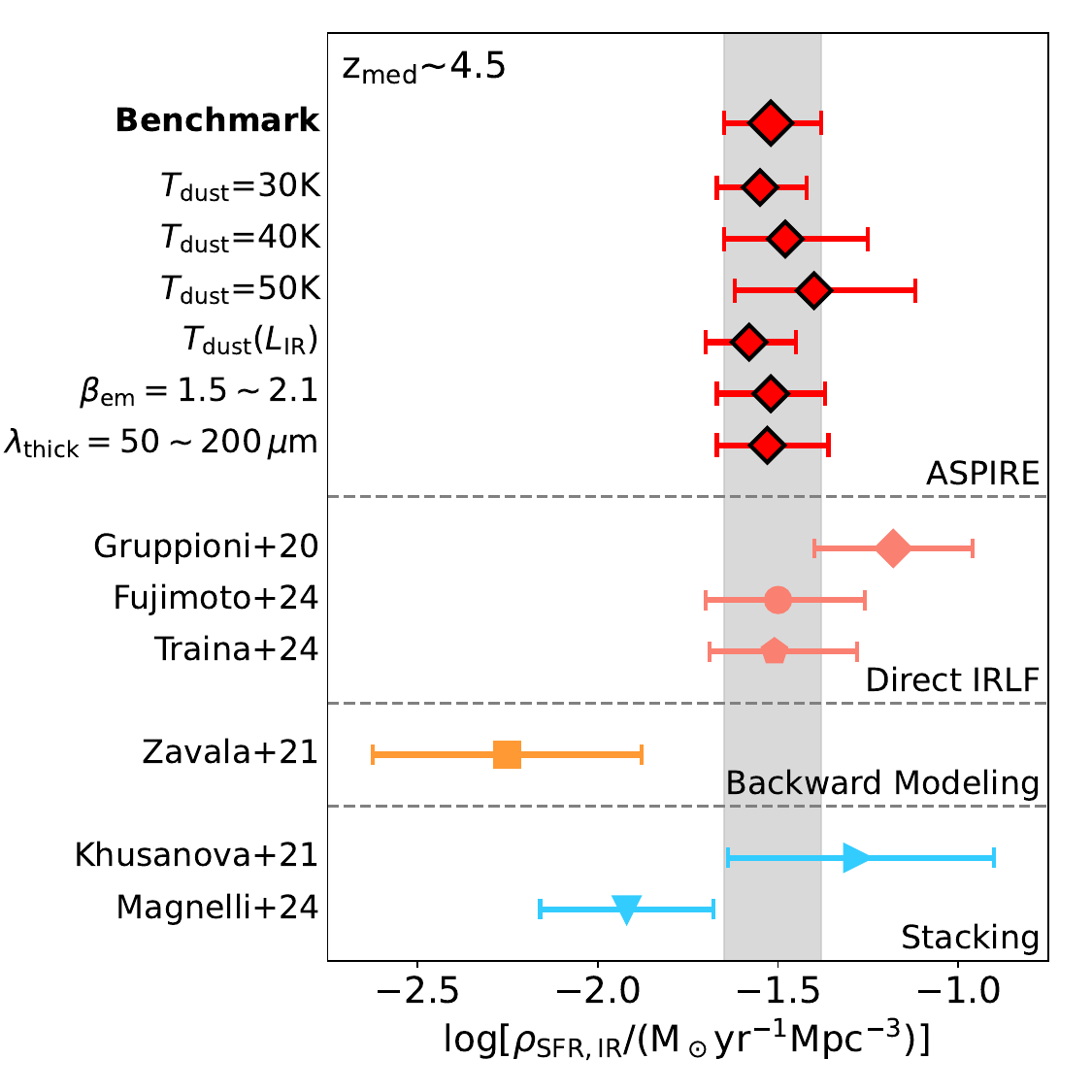}
\caption{16-50-84\% credible-interval constraints of obscured SFRD at $z \sim 4.5$.
Results from ASPIRE under various far-IR SED and dust temperature assumptions are shown on the top as solid red diamonds.
Literature results through direct measurement and integral of IRLF are shown as salmon-pink symbols \citep[][interpolated to $z\sim4.5$]{gruppioni20, fujimoto24, traina24}.
The result from \citet{zavala21} through backward evolutionary modeling is shown as the orange square.
Results based on stacking of ALMA data in bins of stellar masses or UV luminosities are shown in cyan triangles \citep{khusanova21,magnelli24}. 
The vertical gray band shows the 16--84\% credible-interval of the ``benchmark'' obscured SFRD measured with ASPIRE.
}
\label{fig:bench}
\end{figure}

\subsection{Far-IR SED}
\label{ss:05a_fir}

In order to determine IRLF and SFRD, we make use of energy-balance SED fitting code \textsc{cigale} to derive \lir\ (Section~\ref{ss:03b_sed}).
The accuracy of \lir\ thus rely on the validity of the energy-balance and far-IR SED assumptions.

It is known that DSFGs may not obey the energy-balance assumption along the observed sightline, simply because of patchy dust geometry.
Previous studies have shown that DSFGs can have relatively blue UV continuum slopes, placing them above the $\mathrm{IRX} - \beta_\mathrm{UV}$ relation for normal star-forming galaxies \citep[e.g.,][]{penner12, oteo13, casey14b}.
One of the most recent examples is HDF850.1 at $z=5.18$, which is $\sim$\,100 times above the $\mathrm{IRX} - \beta_\mathrm{UV}$ relation because of the leakage of UV photons \citep{sun24}. 
Similar effect has been seen for at least one DSFG in the ASPIRE sample (J0244m5008.C03), and the best-fit SED model does not reproduce the flux density in the F115W band satisfactorily (Figure~\ref{fig:sed}).

To quantify the potential uncertainty of obscured SFRD determination with the energy balance assumption (hereafter the ``benchmark'' SFRD), we derive IRLF and obscured SFRD following the same methods presented in Section~\ref{sec:04_res} but assuming different dust temperatures.

First of all, the \textsc{cigale} best-fit \lir\ used by ``benchmark'' SFRD is roughly equivalent to that computed from modified black body SED with $T_\mathrm{dust} = 33$\,K and $\beta_\mathrm{em} = 1.8$.
This is similar to the typical $T_\mathrm{dust}$ for galaxies with $L_\mathrm{IR}=10^{12}$\,\lsun\ observed out to $z\sim4.5$ (\citealt{sun22a}; also modeled with MBB SED with $\beta_\mathrm{em} = 1.8$), in which the direct $T_\mathrm{dust}$ measurements were obtained through Herschel/SPIRE at 250--500\,\,micron thanks to lensing magnification.

However, arguably there is no consensus of the typical $T_\mathrm{dust}$ for LIRGs at $z\sim5$.
On one hand, studies have suggested that \mstar-selected galaxies have higher dust temperatures at higher redshifts through the stacking of Herschel data (\citealt{magnelli14,schreiber18b}; caution shall be taken for the large beam size up to $\sim$35\arcsec).
Higher dust temperatures at higher redshifts is a natural consequence of warmer cosmic microwave background (CMB) heating.
On the other, direct $T_\mathrm{dust}$ measurements of \lir-selected DSFGs also suggest no or weak redshift evolution of $T_\mathrm{dust}$ up to $z\sim4.5$ \citep[e.g.,][]{dudzevic20,drew22,sun22a}, and therefore the $T_\mathrm{dust}(z)$ evolution at fixed stellar mass can simply be a combined effect of the weak evolution of the $L_\mathrm{IR} – T_\mathrm{dust}$ relation and strong evolution of the so-called star-forming ``main sequence''.

Nevertheless, we explore the possible $T_\mathrm{dust}$ range at 30 to 50\,K, derive the \lir\ for ASPIRE sources through these $T_\mathrm{dust}$ assumptions with \citealt{casey12} SED (MBB $\beta_\mathrm{em}=1.8$ and mid-IR $\alpha_\mathrm{MIR} = 2.0$) and plot the resultant obscured SFRD measurements in Figure~\ref{fig:bench}.
With hotter $T_\mathrm{dust}$, the resultant obscured SFRD is larger because $L_\mathrm{IR}$ is proportional to $T_\mathrm{dust}^4$ (the Stefan–Boltzmann law).
However, the increase of obscured SFRD is only 0.12\,dex (despite with larger uncertainty) with $T_\mathrm{dust} = 50$\,K assumption compared with the ``benchmark''. 
This is a consequence of flattened IRLF at the faint end.
Note that we also consider CMB heating effect following \citet{dacunha13}, while the effect is found to be small at $z\sim5$ where CMB temperature $T_\mathrm{CMB}$ is 16\,K (\lir\ boost is $\lesssim$\,0.01\,dex).

We also consider the very hot IR SED template of Haro 11, a local moderately low-metallicity galaxy that undergoes starburst.
\citet{derossi18} suggested that the IR SED of Haro 11 could resemble those of luminous DSFGs at $z=5-7$, many of which are quasar host galaxies \citep{lyu17} or selected through single-dish telescopes including SPT \citep[e.g.,][]{strandet16}.
With the Haro 11 SED assumption, all galaxies in our sample exhibit \lir\ of ULIRGs, and thus we will not be able to constrain the faint-end slope of IRLF any more, and literature measurements will dominate the IRLF fitting.
The resultant obscured SFRD is similar to that approximated by $T_\mathrm{dust}=50$\,K (Figure~\ref{fig:bench}).

We also derive \lir\ using the $\lambda_\mathrm{peak}(L_\mathrm{IR})$ relation in \citet{casey18b}, which is equivalent to:
\begin{equation}
T_\mathrm{dust} / \si{K} = 32.7\, [L_\mathrm{IR}/(10^{12}\,\si{L_\odot})]^{0.09}
\end{equation}
under the adopted SED assumption, and the resultant SFRD is slightly lower than our ``benchmark'' by 0.05\,dex because of the relatively low \lir\ of ASPIRE DSFGs (mostly LIRGs).

We also experiment different assumptions of dust emissivity and rest-frame wavelength where the dust optical depth is unity, both of which were fixed in previous modeling ($\beta_\mathrm{em} = 1.8$ and $\lambda_\mathrm{thick} = 100$\,\micron).
We find that the change of $\beta_\mathrm{em}$ by $\pm0.3$ will result in a variation of 0.02\,dex for $\rho_\mathrm{SFR,IR}$, and the change of $\lambda_\mathrm{thick}$ by a factor of 2 will result in a variation of 0.07\,dex.
\textred{These variations are relatively insignificant compared with the uncertainty of SFRD.}

\textred{Finally, we also consider the potential AGN contribution to the 1.2-mm ALMA continuum and thus far-IR SED. 
Given that we have ruled out the existence of bright point-like structure in NIRCam imaging ($>$\,27.6\,AB mag at F356W, 10$\sigma$ limit), any AGNs must be dust-obscured if they exist in ASPIRE DSFGs. 
We make use of the AGN SED templates constructed by \citet{lyu17} and \citet{lyu18}.
At the F356W detection limit, AGN at $A_V<3$ cannot contribute to more than 25\% of observed ALMA flux density, and thus our SFRD measurement remains valid. 
However, we cannot rule out the existence of AGN with larger $A_V$ that may contribute to the 1.2\,mm flux density more significantly \citep[e.g.,][]{symeonidis16,mckinney21}, although obscured AGNs are not required to interpret the observational characteristics of ASPIRE DSFGs.
Obscured AGN contribution to the IR SEDs of $z=4-6$ (U)LIRGs will need to be quantified through future JWST/MIRI observations and far-IR missions.
}

In conclusion, we argue that our determination of obscured SFRD is relatively robust against largely unknown far-IR SEDs, which are only loosely constrained by ALMA photometry in just one band. 
The ASPIRE determinations of obscured SFRD, regardless of $T_\mathrm{dust}$ assumptions, are all higher than that of \citet{zavala21} but consistent with recent direct measurements from \citet{fujimoto24} and \citet{traina24}, and exhibit smaller uncertainties than literature measurements relying on stacked ALMA data \citep{khusanova21,magnelli24}.
To further improve the accuracy of obscured SFRD determination, direct $T_\mathrm{dust}$ measurements through high-frequency ALMA continuum observations of $z>4$ DSFGs would be very helpful \citep[e.g.,][]{faisst20,mitsuhashi24,valentino24,villanueva24}.

\begin{figure}[!t]
\centering
\includegraphics[width=\linewidth]{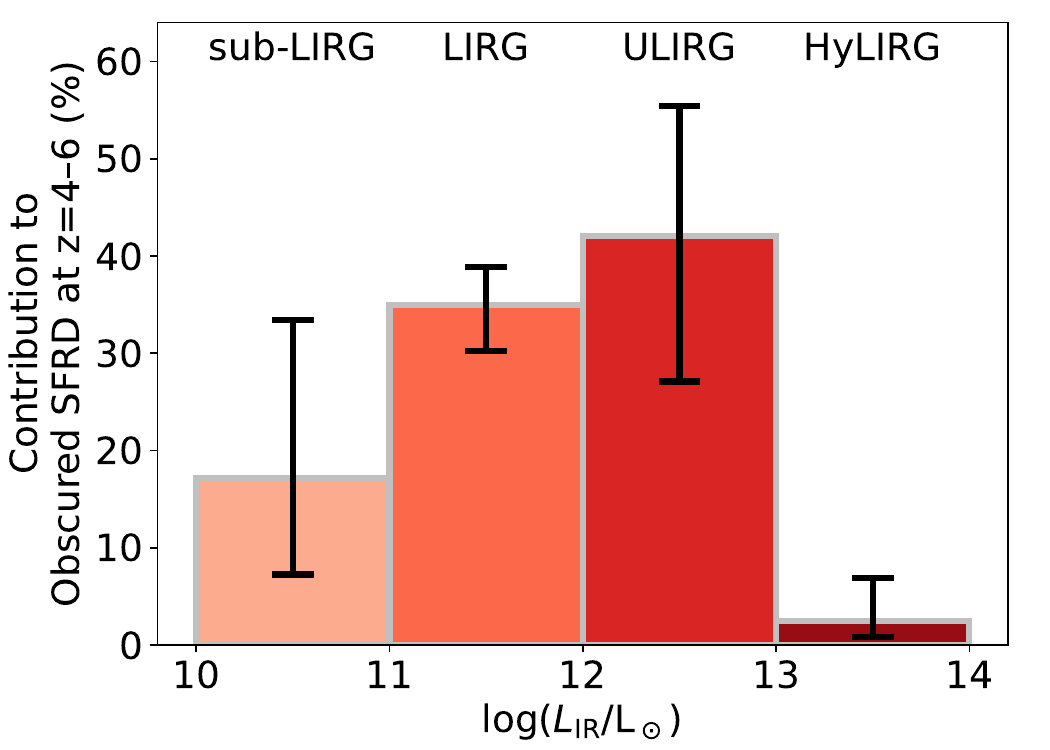}
\caption{The contribution to obscured cosmic SFRD at $z=4-6$ from sub-LIRGs, LIRGs, ULIRGs and HyLIRGs at \lir\ from $10^{10}-10^{14}$\,\lsun.
LIRGs and ULIRGs at $L_\mathrm{IR} = 10^{11} - 10^{13}$\,\lsun\ contribute to the majority ($81_{-21}^{+11}$\%) of obscured SFRD at this epoch. 
}
\label{fig:contri}
\end{figure}

\subsection{The IRLF at the Bright End}
\label{ss:05b_bend}

Because of limited survey area, our ASPIRE DSFG sample cannot constrain the IRLF at the bright end (i.e., $L_\mathrm{IR} > L_\mathrm{IR}^{\star} \sim 10^{12.6}$\,\lsun) in a self-contained manner.
Therefore, we have to include literature measurements in our fitting, which could lead to the concern of \zsp\ completeness and \zph\ validity.

Figure~\ref{fig:contri} shows the contribution to the obscured cosmic SFRD at $z=4-6$ by DSFGs in four \lir\ bins, namely sub-LIRGs ($10^{10}\,\lsun \leq L_\mathrm{IR} < 10^{11}\,\lsun$), LIRGs, ULIRGs and HyLIRGs ($10^{13}\,\lsun \leq L_\mathrm{IR} < 10^{14}\,\lsun$).
It is evident that both LIRGs and ULIRGs contribute to the majority ($81_{-21}^{+11}$\%) of the obscured SFRD at this epoch.
However, the contribution from ULIRGs is subject to a relatively large uncertainty ($44\pm14$\%) because of the uncertainty of IRLF at the bright end.

Although the spectroscopic completeness of literature measurement could be a concern, we argue that the derived $z=4-6$ IRLF in Section~\ref{fig:IRLF} is actually comparable to recent spectroscopic evidence.
Simply considering GN10 ($z=5.30$; \citealt{riechers20}) and HDF850.1 ($z=5.18$; \citealt{walter12}) whose \ha\ lines are detected through FRESCO NIRCam WFSS in the F444W band (\citealt{sun24}; see also \citealt{hd24,xiao24}), we can roughly estimate a volume density of $\Phi = 10^{-5.0\pm0.4}$\,\si{Mpc^{-3}.dex^{-1}} (cosmic variance considered) within the luminosity bin $10^{12.5}-10^{13.1}$\,\lsun\ assuming the FRESCO survey volume at $z=4.9-6.0$.
This is consistent with the volume density expected from our best-fit IRLF at the bin center ($\Phi = 10^{-5.4}$\,\si{Mpc^{-3}.dex^{-1}}).
However, we also note that both galaxies reside in a complex overdense environment \citep{calvi21, calvi23, hd24, sun24}, and there could be more galaxies in both GOODS field that could contribute to the volume density \citep{xiao24}.

In JWST Cycle-3, there are four planned large / treasury NIRCam WFSS surveys over much larger volumes, including NEXUS (PID: 5105, PI: Shen, Y.; \citealt{sheny24}), COSMOS-3D (PID: 5893, PI: Kakiichi, K.), POPPIES (PID: 5398, PI: Kartaltepe, J.), SAPPHIRES (PID: 6434, PI: Egami, E.), totaling more than one thousand hours of JWST time. 
These observations, in synergy with obtained / planned (sub-)millimeter surveys in cosmological deep fields (COSMOS, including \citealt{casey13, casey21, geach17, wangw17, simpson19}; and NEP, including \citealt{shim20,hyun23}), will provide powerful and essential spectroscopy of high-redshift DSFGs at the bright end of IRLF, constraining the obscured cosmic star-formation history towards the Epoch of Reionization with unprecedented accuracy.

\section{Summary}
\label{sec:06_sum}

In this paper, we present a stringent measurement of the dust-obscured SFRD at $z=4-6$ through the ASPIRE JWST and ALMA survey.
ASPIRE obtained JWST NIRCam WFSS survey at 3.1--4.0\,\micron\ and ALMA mosaics at 1.2\,mm along 25 quasar sightlines. 
We make use of this JWST and ALMA dataset to identify DSFGs at $z_\mathrm{spec} = 4 - 6$ through ALMA continuum detections and JWST spectroscopy.
The main results are summarized as follows:

\begin{enumerate}

\item We identify eight DSFGs at $z_\mathrm{spec}=4-6$ through the detections of \ha\ (for seven sources at $z=4.07-4.80$) or \oiii\,$\lambda$5008 (for one source at $z=5.55$) lines with NIRCam WFSS.
These DSFGs are discovered in seven out of 25 ASPIRE quasar fields totaling a survey area of $\sim$\,35\,arcmin$^2$.
For six DSFGs, we also detect at least one more faint lines such as \hb, \oiii\,$\lambda$4960, \nii\,$\lambda$6585 and \sii\,$\lambda\lambda$6718,6733.

\item We conduct energy-balance SED modeling of these DSFGs through JWST and ALMA photometry (four bands in total).
The median stellar mass is $\log(M_\mathrm{star}/\si{M_\odot}) = 10.3 \pm 0.2$, and the median IR luminosity is $\log(L_\mathrm{IR} / \si{L_\odot}) = 11.7\pm0.1$.

\item 96$\pm$2\% of the star formation in $z\sim5$ DSFGs probed by ASPIRE are obscured by dust.
However, the escape fractions of \ha\ photon is anti-correlated with their IR luminosities, resulting in an almost constant \ha\ luminosity across 2-dex wide span of \lir, which is also well above the $5\sigma$ detection limit with NIRCam WFSS.
We observe a typical $\tau_\mathrm{UV} / \tau_\mathrm{H\alpha}$ ratio of $1.61\pm0.34$ for $z\sim5$ DSFGs, slightly higher than that of local starburst, which could lead to an enhanced \ha\ visibility in these $z\sim5$ DSFGs.

\item We obtain 1.2-mm number count of DSFGs at $z=4-6$, finding it higher than previous determination based on ALMA Cycle-4 large program ASPECS \citep{popping20,zavala21} but similar to that from ALMA Cycle-5 large program ALPINE \citep{gruppioni20}.
We argue that our measurement is robust against cosmic variance (shot noise) because of averaging 25 independent sightlines, while the ASPECS survey in the HUDF region encounters a void of DSFGs at $z=4-6$ because of strong shot noise.

\item We obtain spectroscopically complete measurement of the infrared luminosity function at $z=4-6$ down to $L_\mathrm{IR} \sim 2 \times 10^{11}$\,\lsun.
We observe a flattening of the faint-end slope ($\alpha = 0.59_{-0.45}^{+0.39}$) of the IRLF.
Our measurements are consistent with previous measurements including \citet{gruppioni20}, \citet{fujimoto24} and \citet{traina24} at the faint end (with spectroscopic incompleteness), but slightly lower than a few model predictions.

\item We derive an obscured SFRD of $\log[\rho_\mathrm{SFR,IR}/(\si{M_\odot.yr^{-1}.Mpc^{-3}})] = -1.52_{-0.13}^{+0.14}$ at $z=4-6$ (median $z = 4.5$).
This is comparable to the total (obscured and unobscured) SFRD modeled by \citet{md14}, suggesting that the majority ($66_{-7}^{+7}$\%) of the total SFRD is obscured by dust at this epoch.
Our measurement is similar to that of direct measurements with recent literature \citep[e.g.,][]{fujimoto24,traina24}, but $\sim$\,5 times higher than that of \citet{zavala21} through a backward evolutionary modeling.

\item We conclude that our measurement of SFRD is relatively robust against uncertainty in far-IR SED shape (primarily dust temperature).
Future ALMA high-frequency observations will be helpful to determine the dust temperature and improve the accuracy of obscured SFRD.
Although our result suggests the majority ($81_{-21}^{+11}$\%) of obscured SFRD is contributed by both LIRGs and ULIRGs, the current uncertainty at the bright end of IRLF (especially \zsp\ incompleteness) will need to be resolved with planned JWST NIRCam WFSS surveys over much larger volume.

\end{enumerate}

\section*{Acknowledgment}

We thank JWST \#2078 program coordinator Weston Eck and ALMA program \#2022.1.01077.L contact scientist Catherine Vlahakis for their supports to our observing programs.
We thank NRAO for providing computational resource supports on ALMA data processing. 
\textred{We thank the anonymous referee for their helpful comments. 
F.S.\ thanks Seiji Fujimoto for helpful discussions.}

F.S.\ acknowledges funding from JWST/NIRCam contract to the University of Arizona, NAS5-02105 and support for JWST program \#2883 provided by NASA through a grant from the Space Telescope Science Institute, which is operated by the Association of Universities for Research in Astronomy, Inc., under NASA contract NAS 5-03127.
F.W.\ and X.F.\ acknowledge support from NSF Grant AST-2308258.
J.B.C.\ acknowledges funding from the JWST Arizona/Steward Postdoc in Early galaxies and Reionization (JASPER) Scholar contract at the University of Arizona. 
L.C.\ acknowledges support by grant PIB2021-127718NB-100 from the Spanish Ministry of Science and Innovation/State Agency of Research MCIN/AEI/10.13039/501100011033 and by “ERDF A way of making Europe”.
Y.K.\ thanks the support of the German Space Agency (DLR) through the program LEGACY 50OR2303.
R.A.M.\ acknowledges support from the Swiss National Science Foundation (SNSF) through project grant 200020\_207349.
S.Z.\ acknowledges support from the National Science Foundation of China (no.\ 12303011).

This paper makes use of the following ALMA data: ADS/JAO.ALMA\#2022.1.01077.L. ALMA is a partnership of ESO (representing its member states), NSF (USA) and NINS (Japan), together with NRC (Canada), MOST and ASIAA (Taiwan), and KASI (Republic of Korea), in cooperation with the Republic of Chile. The Joint ALMA Observatory is operated by ESO, AUI/NRAO and NAOJ. The National Radio Astronomy Observatory is a facility of the National Science Foundation operated under cooperative agreement by Associated Universities, Inc.

This work is based on observations made with the NASA/ESA/CSA James Webb Space Telescope. The data were obtained from the Mikulski Archive for Space Telescopes at the Space Telescope Science Institute, which is operated by the Association of Universities for Research in Astronomy, Inc., under NASA contract NAS 5-03127 for JWST. These observations are associated with program \#2078.
The JWST data presented in this paper were obtained from the Mikulski Archive for Space Telescopes (MAST) at the Space Telescope Science Institute. 
The specific observations analyzed can be accessed via \dataset[10.17909/vt74-kd84]{https://doi.org/10.17909/vt74-kd84}.
Support for program \#2078 was provided by NASA through a grant from the Space Telescope Science Institute, which is operated by the Association of Universities for Research in Astronomy, Inc., under NASA contract NAS 5-03127.


\vspace{5mm}
\facilities{JWST(NIRCam), ALMA}


\software{\textsc{astropy} \citep{2013A&A...558A..33A,2018AJ....156..123A}, 
\textsc{casa} \citep{casa07,casa22},
\textsc{cigale} \citep{cigale09,cigale19},
\textsc{emcee} \citep{emcee}, 
\textsc{galfit} \citep{galfit}, 
\textsc{jwst} \citep{bushouse_jwst},
\textsc{lenstool} \citep{lenstool},
\textsc{photutils} \citep{photutils},
\textsc{webbpsf} \citep{webbpsf},
}



\setcounter{figure}{0}
\renewcommand{\thefigure}{\thesection\arabic{figure}}

\appendix

\section{Lensing Magnification}
\label{apd:01_lens}

The highest-redshift source in our sample, J0109m3047.C02 ($z=5.549$) is very close to a foreground galaxy (separation $\sim$\,0\farcs8).
J0109m3047.C02 is also extended in morphology along the tangential direction to this foreground galaxy, indicating that J0109m3047.C02 is strongly lensed by the foreground source.
To correctly account for the lensing effect in the analyses of its physical properties and lensing bias in the determination of LF, we model the lensing magnification through the following steps.

We first obtain optical and near-IR photometry of the foreground lens. 
At optical wavelength, the foreground galaxy is much brighter than J0109m3047.C02 and the other bluer source to the northwest as seen in Figure~\ref{fig:cutout}. Therefore, we directly make use of DES DR2 \citep{desdr2} photometry in $g$, $r$, $i$, $z$ and $Y$ band.
At JWST/NIRCam wavelength, we carefully model the source morphology with \textsc{galfit} \citep{galfit}.
Because J0109m3047.C02 appears clumpy in the image, we include ten components of S\'ersic profiles or point-spread functions (PSFs) to model the complex image in F115W, F200W and F356W band.
We make use of PSF model generated through \textsc{webbpsf} \citep{webbpsf}.
Except for the foreground lens component, the best-fit morphological models of other components are subtracted from the mosaicked images. 
We therefore obtain the near-IR aperture photometry of the foreground lens on the residual images (22.90$\pm$0.05, 21.82$\pm$0.04, 21.12$\pm$0.05\,AB mag in F115W, F200W, F356W band, respectively).

Because no obvious emission or absorption line could be identified in the grism spectra of the foreground lens, we derive the photometric redshifts through SED modeling.
We feed the photometric measurements into \textsc{CIGALE} \citep{cigale19} with redshift grids from $z=0.5$ to 2.5 ($\Delta z$\,=\,0.05), and derive the reduced $\chi_\mathrm{red}^2(z)$ and stellar mass \mstar$(z)$ for each of the fitting.
We derive the probability distribution of redshifts through $P(z) \propto \exp[- \chi_\mathrm{red}^2(z) / 2]$ and normalization $\int_{0.5}^{2.5} P(z) dz = 1$, and conclude a photometric redshift of \zph\,=\,1.18$\pm$0.05 for the foreground lens.
At this redshift no strong hydrogen line is expected in the F356W grism spectrum, consistent with the aforementioned non-detection.
We randomly draw redshifts from the $P(z)$ distribution, and then draw the corresponding \mstar\ from the best-fit \mstar$(z)$ and the uncertainties.
Assuming the stellar-mass Tully-Fisher Relation measured at $z\sim1$ \citep{ubler17}, we derive velocity dispersion ($\sigma_v$) for each of the drawn pairs of $z$ and \mstar.
The median and $1\sigma$ uncertainty of velocity dispersion is $\sigma_v = 150\pm28$\,\si{km.s^{-1}} for the foreground lens.

We then model the foreground lens using the drawn redshifts and $\sigma_v$ through software \textsc{lenstool} \citep{lenstool}.
We assume a singular isothermal spherical distribution of matter in the foreground potential.
We therefore derive a lensing magnification of $\mu = 1.91 \pm 0.50$ for the ALMA continuum source.
However, we notice that differential lensing effect is seen for J0109m3047.C02 because its complex morphology at different wavelengths.
The light-weighted (through image segments) magnification for J0109m3047.C02 in F115W, F200W and F356W band is $1.58 \pm 0.33$, $1.54 \pm 0.29$ and $ 1.53 \pm 0.28$, respectively.
These magnification factors have been corrected when we infer the physical properties (e.g., luminosity and mass) of J0109m3047.C02.

\section{Completeness Correction}
\label{apd:02_comp}

We simulate ALMA observations with \textsc{casa} to derive the completeness correction at each ALMA source flux and S/N.
First of all, we assume a 2D circular Gaussian profile for ALMA continuum source with an effective radius $R_\mathrm{e} = 0\farcs23$ (i.e., 1.5\,kpc at $z=4.5$). 
This is consistent with the sizes that we measured from image-plane fitting (with \textsc{casa} \texttt{imfit}) for ASPIRE DSFGs detected at $z=4-6$ and $\mathrm{S/N} > 10$.
We scale the surface brightness profile such that the total flux density is 0.10--1.0\,mJy at 1.2\,mm in each of our run of simulations.
ALMA continuum observation measurement sets are then simulated through \textsc{casa} \texttt{simobserve} command with the same configuration and similar observing condition (exposure time and precipitated water vapor) as those of actual ASPIRE observations.
With each input flux density, we simulate the observations for 100 times, produce the continuum images at both native and 1\farcs0-tapered resolution with the same pipeline adopted for real data.
We then identify peaks in simulated images within 1\arcsec\ from the injected source position. 
The detection criteria are the same as those adopted in Section~\ref{ss:02b_alma}.

\begin{figure}
\centering
\includegraphics[width=0.5\linewidth]{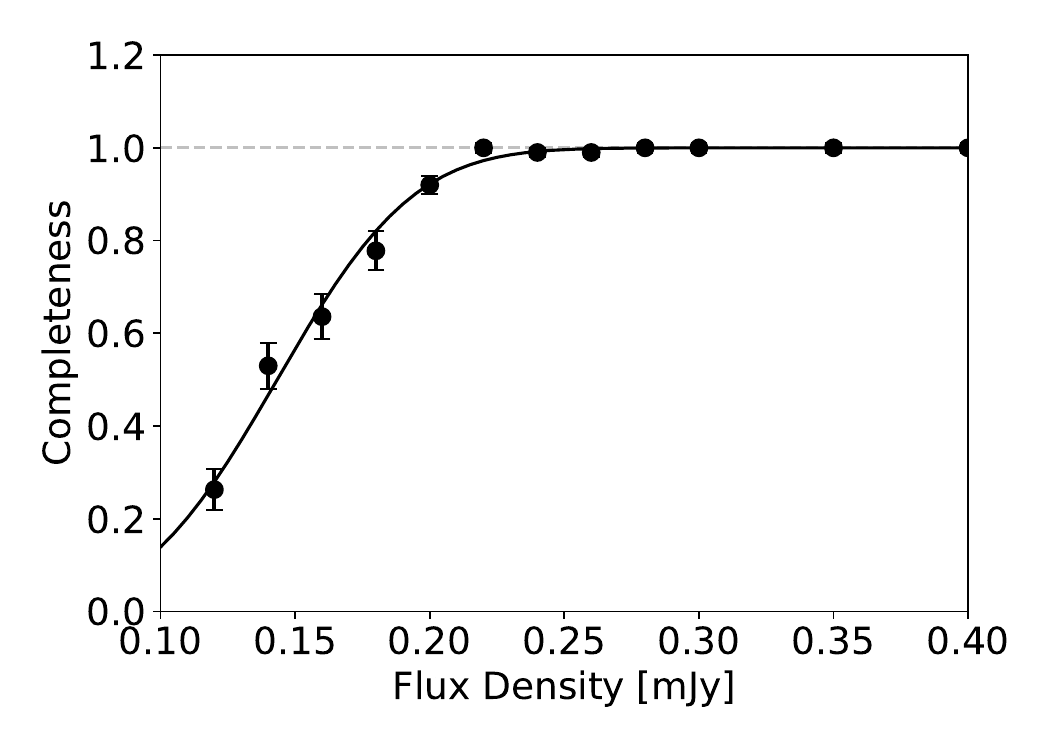}
\vspace{-12pt}
\caption{Completeness as a function of ALMA flux density on 1\farcs0 \textit{uv}-tapered images (before primary beam response correction).
Best-fit error function is shown as the solid black line.
}
\label{fig:complete}
\end{figure}

At each injected source flux density, we compute the fraction of sources in simulated images that are above our detection threshold.
Such a fraction is therefore considered as the survey completeness at the injected source flux, and Figure~\ref{fig:complete} shows the completeness as a function of ALMA flux density on the tapered images.
We also conduct statistics on native-resolution images, but the results are less relevant because all of the ASPIRE DSFGs $z=4-6$ exhibit higher S/N on tapered images than native images (and thus completeness is higher).
We fit the measured completeness with an error function:

\begin{equation}
    C(f) = \frac{1}{2}  \mathrm{erf}\big(\frac{S - S_0}{\sigma}\big) + \frac{1}{2}
\end{equation}
and we find that the completeness is 50\%\ at source flux density $S = S_0 = 0.143\pm0.003$\,mJy.
All ASPIRE DSFGs at $z=4-6$ show high completeness ($C\sim1$) because of their high fidelity of detections ($S_\mathrm{1.2mm} \geq 0.24$\,mJy and $\mathrm{S/N} \geq 6.5$). 
We also rewrite the completeness function in terms of ALMA source S/N and use it for the IRLF completeness calculation.

Through our injection and detection experiment, we also verify the 1.1$\times$ scaling factor applied to the peak flux density measured on tapered image ($f_\mathrm{tap}$; Section~\ref{ss:03a_spec}).
We also conclude negligible flux boosting effect (e.g., Eddington bias; $<$3\% through our simulations) in the S/N and flux density range of our targets. 


\bibliography{00_main}{}
\bibliographystyle{aasjournal}



\suppressAffiliationsfalse
\allauthors

\end{document}